\documentclass[pre,twocolumn,showpacs,preprintnumbers,superscriptaddress]{revtex4}

\usepackage{amsmath}
\usepackage{graphicx}
\usepackage{mathrsfs}
\usepackage{amsmath}
\usepackage{amssymb}
\usepackage{graphicx}
\usepackage[british]{babel}

\begin{document}

\preprint{\today.}

\title{Effective surface motion on a reactive cylinder of particles that 
perform intermittent bulk diffusion}

\author{Aleksei V. Chechkin}
\affiliation{Institute for Theoretical Physics NSC KIPT,
Akademicheskaya st.1, 61108 Kharkov, Ukraine}
\affiliation{School of Chemistry, Tel Aviv University, 69978 Tel Aviv,
Israel}
\author{Irwin M. Zaid}
\affiliation{Physics Department, Technical University of Munich,
85747 Garching, Germany}
\author{Michael A. Lomholt}
\affiliation{MEMPHYS - Center for Biomembrane Physics, Department of
Physics and Chemistry, University of Southern Denmark, Campusvej 55,
5230 Odense M, Denmark}
\author{Igor M. Sokolov}
\affiliation{Institut f{\"u}r Physik, Humboldt Universit{\"a}t
zu Berlin, Newtonstra{\ss}e 15, 12489 Berlin, FRG}
\author{Ralf Metzler}
\email{metz@ph.tum.de}
\affiliation{Physics Department, Technical University of Munich,
85747 Garching, Germany}
\affiliation{Physics Department, Tampere University of Technology,
FI-33101 Tampere, Finland}


\begin{abstract}
In many biological and small scale technological applications particles
may transiently bind to a cylindrical surface. In between two binding
events the particles diffuse in the bulk, thus producing an effective
translation on the cylinder surface. We here derive the
effective motion on the surface, allowing for additional
diffusion on the cylinder surface itself. We find explicit solutions
for the number of adsorbed particles at one given instant, the effective
surface displacement, as well as the surface propagator. In particular
sub- and superdiffusive regimes are found, as well as an effective
stalling of diffusion visible as a plateau in the mean squared
displacement. We also investigate the corresponding first passage
and first return problems.
\end{abstract}

\maketitle

\section{Introduction}

\begin{figure}
\includegraphics[width=7.2cm]{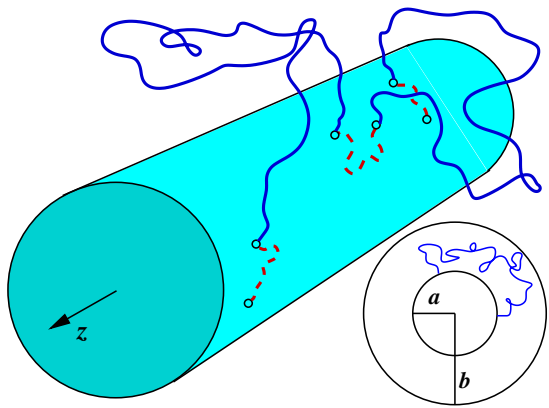}
\caption{A particle diffuses in the bulk (full lines) and intermittently
binds to a cylinder surface on which it may also diffuse (broken lines). This
produces an effective surface motion. We here consider the motion along
the cylinder. Bottom right: frontal of diffusion between inner
and outer cylinder.}
\label{scheme}
\end{figure}

Bulk mediated surface diffusion (BMSD) defines the effective surface motion
of particles, that intermittently adsorb to a surface or diffuse in the
contiguous bulk volume. As sketched in Fig.~\ref{scheme} for a cylindrical
surface, the particle, say, starts on the surface and diffuses along this
surface with diffusion constant $D_s$. Eventually the particle unbinds, and
performs a three-dimensional stochastic motion in the adjacent bulk, before
returning to the surface. Typically, the values of $D_b$ are significantly
larger than $D_s$. The recurrent bulk excursions therefore lead to
decorrelations in the effective surface motion of the particle, and thus to a
more efficient exploration of the surface.

Theoretically
BMSD was previously investigated for a planar surface in terms of scaling
arguments \cite{bychuk,bychuk1}, master equation schemes \cite{revelli},
and simulations \cite{fatkullin}. More recently the first passage problem
between particle unbinding and rebinding for a free cylindrical surface was
derived \cite{levitz}. Following our short communication \cite{rc} we here
present in detail an exact treatment of BMSD for a reactive cylindrical
surface deriving explicit expressions for the surface occupation, the
effective mean squared displacement (MSD) along the surface, and the
returning time distribution from the bulk. In this approach different
dynamic regimes arise naturally from the physical timescales entering our
description. Thus at shorter times we derive the famed superdiffusive
surface spreading with surface MSD of the form $\langle z^2(t)\rangle\sim
t^{3/2}$ \cite{report,igor}
and the associated Cauchy form of the surface probability density
function (PDF). At longer times we obtain an a priori unexpected leveling
off of the surface MSD, representing a tradeoff between an increasing
number of particles that escape into the bulk and the increasing
distance on the surface covered in ever-longer bulk excursions for those
particles that do return to the surface. Only when the system is confined
by an outer cylinder eventually normal surface diffusion will emerge.
Apart from the L{\'e}vy walk-like superdiffusive regime the rich dynamic
behavior found here are characteristic of the cylindrical geometry.

Nuclear magnetic resonance (NMR) measurements of liquids in porous media are
sensitive to the preferred orientation of adsorbate molecules on the local
pore surface, such that surface diffusion on such a non-planar surface
produces spin reorientations and remarkably long correlations times
\cite{kimmich}. Apart from pure surface diffusion the experiment by Stapf
et al. clearly showed the influence of BMSD steps and the ensuing L{\'e}vy
walk-like superdiffusion \cite{stapf}. More recently NMR techniques were
used to unravel the effective surface diffusion on cylindrical mineralic
rods \cite{levitz}, supporting, in particular, the first passage behavior
with its typical logarithmic dependence. BMSD on a cylindrical surface is
also relevant for the transient binding of chemicals to nanotubes
\cite{nanotubes} and for numerous other technological applications
\cite{bychuk1}. In a biological context, BMSD along a cylinder is intimately
related to the diffusive dynamics underlying gene regulation \cite{bvh,berg}:
DNA binding proteins diffuse not only in the bulk but intermittently bind
non-specifically to the DNA, approximately a cylinder, and perform a
one-dimensional motion along the DNA chain, as proved experimentally
\cite{wang,bonnet}. The interplay between bulk and effective
surface motion improves significantly the search process of the protein for
its specific binding site on the DNA. Similarly the net motion of motor
proteins along cytoskeletal filaments is also affected by bulk mediation.
Namely, the motors can fall off the cellular tracks and then rebind to the
filament after a bulk excursion \cite{motor}. Outside of biological cells the
exchange behavior between cell surface and surrounding bulk is influenced by
bulk excursions, the cylindrical geometry being of relevance for a large class
of rod-shaped bacteria (bacilli) and their linear arrangements \cite{bacillus}.

The dynamics revealed by our approach may also be important for the
quantitative understanding of colonialization processes on surfaces in
aqueous environments when convection is negligible: suppose that bacteria
stemming from a localized source, for instance, near to a submarine hot
vent, start to grow on an offshore pipeline. From this mother colony new
bacteria will be budding and enter the contiguous water. The L{\'e}vy
dust-like distribution due to BMSD will then make sure that bacteria can
start a new colony, that is disconnected from the former, and therefore
give rise to a much more efficient spreading dynamics over the pipeline.

In all these examples it is irrelevant which specific trajectory the particles
follow in the bulk, the interesting part is the effective motion on the
cylinder surface. We here analyze in detail this bulk mediated surface
diffusion on a long cylinder.

\section{Characteristic time scales and important results}
\label{important}

In this Section we introduce the relevant time scales of the problem of bulk
mediated surface diffusion of the cylindrical geometry presented in
Fig.~\ref{scheme} and collect the most important results characteristic for
the effective surface motion of particles. As we are interested only in the
motion along the cylinder axis $z$, we consider the rotationally symmetric
problem with respect to the polar angle $\theta$, that will therefore not
appear explicitly in the following expressions (compare also
Sec.~\ref{results}).
Since the full analytical treatment of the problem involves tedious
calculations we first give an overview of the most important results, leaving
the derivations to the forthcoming Sections and Appendices.

\subsection{Characteristic time scales}

Using the result (\ref{nsol_fl}) for the Fourier-Laplace transform $n(k,s)$
of the density of particles on the cylinder surface, we compute
the Laplace transform of the number of surface particles,
\begin{eqnarray}
\nonumber
N_s(s)&=&\int_{-\infty}^{\infty}n(z,s)dz=n(k,s)\Big|_{k=0}\\
&=&\frac{N_0}{\displaystyle
s+\kappa\sqrt{\frac{s}{D_b}}\frac{\Delta_1(0,s)}{\Delta(0,s)}},
\label{ns_l}
\end{eqnarray}
where $\kappa$ is a surface-bulk coupling constant defined below, $D_b$ is the
bulk diffusion constant,
\begin{eqnarray}
\nonumber
\Delta_1(0,s)&=&K_1(a\xi)I_1(b\xi)-I_1(a\xi)K_1(b\xi),\\
\Delta(0,s)&=&I_0(a\xi)K_1(b\xi)+K_0(a\xi)I_1(b\xi),
\end{eqnarray}
and $\xi\equiv\sqrt{s/D_b}$. The $I_{\nu}$ and $K_{\nu}$ denote modified
Bessel functions. We define the Laplace and Fourier transforms
of the surface density $n(z,t)$ through
\begin{equation}
n(k,t)=\mathscr{F}\{n(z,t)\}=\int_{-\infty}^{\infty}e^{ikz}n(z,t)dz
\end{equation}
and
\begin{equation}
n(z,s)=\mathscr{L}\{n(z,t)\}=\int_0^{\infty}e^{-st}n(z,t)dt.
\end{equation}
Here and in the following we express the transform of a function by explicit
dependence on the Fourier or Laplace variable, thus, $n(k,s)$ is the
Fourier-Laplace transform of $n(z,t)$.

From expression (\ref{ns_l}) we recognize that in the limit $\kappa\to0$
the number of particles on the cylinder surface does not change, i.e.,
$N(t)=N_0$. The coupling parameter according to Eq.~(\ref{kappa}) is
connected to the unbinding time scale $\tau_{\mathrm{off}}$, the bulk
diffusivity $D_b$, and the binding rate $k_b$ through $\kappa=D_b/[k_b
\tau_{\mathrm{off}}]$. Vanishing $\kappa$ therefore corresponds to an
infinite time scale for unbinding. This observation allows us to introduce
a characteristic coupling time
\begin{equation}
t_{\kappa}\equiv\frac{D_b}{\kappa^2}=\frac{k_b^2\tau_{\mathrm{off}}^2}{D_b}.
\end{equation}
Note that the binding constant $k_b$ has dimension $\mathrm{cm}/\mathrm{sec}$,
see Section \ref{results}. As in the governing equations the coupling constant
$\kappa$ and the bulk diffusivity $D_b$ are the relevant parameters, the
characteristic time $t_{\kappa}$ in a scaling sense is uniquely defined.
When the coupling between bulk and cylinder surface is weak, $\kappa\to0$,
the corresponding coupling time $t_{\kappa}$ diverges. It vanishes when the
coupling is strong, $\kappa\to\infty$.

While the time scale $t_{\kappa}$ is characteristic of the bulk-surface
exchange, the geometry of the problem imposes two additional characteristic
times. Namely, the inner and outer cylinder radii involve the time scales
\begin{equation}
\label{t_a}
t_a\equiv\frac{a^2}{D_b}
\end{equation}
and
\begin{equation}
\label{t_b}
t_b\equiv\frac{b^2}{D_b},
\end{equation}
respectively. By definition, $t_b$ is always larger than $t_a$.
For times shorter than the scale $t_a$ a diffusing particle
behaves as if it were facing a flat surface, while for times longer then
$t_a$ it can sense the cylindrical shape of the surface. Similarly, $t_b$
defines the scale when a particle starts to engage with the outer cylinder
and therefore senses the confinement. With the help of these time scales we
can rewrite expression (\ref{ns_l}) for the number of surface particles
in the form
\begin{equation}
N_s(s)=\frac{N_0t_{\kappa}^{1/2}}{s^{1/2}\Big((st_{\kappa})^{1/2}+\Delta_1
(0,s)/\Delta(0,s)\Big)},
\end{equation}
where
\begin{equation}
\Delta_1(0,s)=K_1\left(\sqrt{st_a}\right)I_1\left(\sqrt{st_b}\right)-
I_1\left(\sqrt{st_a}\right)K_1\left(\sqrt{st_b}\right)
\end{equation}
and
\begin{equation}
\Delta(0,s)=I_0\left(\sqrt{st_a}\right)K_1\left(\sqrt{st_b}\right)+
K_0\left(\sqrt{st_a}\right)I_1\left(\sqrt{st_b}\right).
\end{equation}

From the characteristic time scales $t_{\kappa}$, $t_a$, and $t_b$ we can
construct the three limits:

(i) Strong coupling limit
\begin{equation}
\label{scl}
t_{\kappa}\ll t_a\ll t_b;
\end{equation}
here the shortest time scale is the coupling time. This regime is the most
interesting as it leads to the transient L{\'e}vy walk-like superdiffusive
behavior.

(ii) Intermediate coupling limit
\begin{equation}
\label{icl}
t_a\ll t_{\kappa}\ll t_b;
\end{equation}
here the superdiffusive regime is considerably shorter, however, an
interesting transition regime is observed.

(iii) Weak coupling limit
\begin{equation}
\label{wcl}
t_a\ll t_b\ll t_{\kappa}.
\end{equation}
To limit the scope of this paper we will not consider this latter case in
the following. However, we note that for $t\ll t_b$ the behavior will be
similar to the $t\ll t_\kappa$ part of the intermediate coupling limit (ii).

\subsection{Important results}

We now discuss the results for the most important quantities characteristic
of the effective surface motion. The dynamic quantities we consider are the
number of particles $N_s(t)$, that are adsorbed to the inner cylinder surface
at given time $t$; as well as the one-particle mean squared displacement
\begin{equation}
\langle z^2(t)\rangle=\frac{1}{N_0}\int_{-\infty}^{\infty}z^2n(z,t)dz.
\end{equation}
This quantity is biased by the fact that an increasing amount of particles
is leaving the surface. To balance for this loss and quantify the effective
surface motion for those particles that actually move on the surface, we
also consider the `normalized' mean squared displacement
\begin{equation}
\langle z^2(t)\rangle_{\mathrm{norm}}=\frac{1}{N_s(t)}\int_{-\infty}^{
\infty}z^2n(z,t)dz.
\end{equation}
The detailed behavior of these quantities will be derived in what follows,
and we will also calculate the effective surface concentration $n(z,t)$
itself. Here we summarize the results for the surface particle number and
the surface mean squared displacements.

\subsubsection{Strong coupling limit}

In Table \ref{tab_strong} we summarize the behavior in the four relevant
time regimes for the case of strong coupling. The evolution of the number
of particles on the surface turns from an initially constant behavior
to an inverse square root decay when the particles engage into surface-bulk
exchange. At longer times, the escape of particles to the bulk becomes
faster and follows a $1/t$ law. Eventually the confinement by the outer
cylinder comes into play, and we reach a stationary limit.

\begin{table*}
\begin{tabular}{|c|c|c|c|}
\hline
& & &\\
\hspace*{0.6cm} Time regime \hspace*{0.6cm} & \hspace*{0.8cm} $N_s(t)$
\hspace*{0.8cm} & \hspace*{2.0cm} $\langle z^2(t)\rangle$  \hspace*{2.0cm} &
\hspace*{1.8cm} $\langle z^2(t)\rangle_{\mathrm{norm}}$ \hspace*{1.8cm}
\\
& & &\\
\hline
& & &\\
$t\ll t_{\kappa}$ & $N_0$ & $\displaystyle 2D_st+\frac{4}{3\sqrt{\pi t_{
\kappa}}}D_bt^{3/2}$ & $\displaystyle 2D_st+\frac{4}{3\sqrt{\pi t_{\kappa}}}
D_bt^{3/2}$\\
& & &\\
\hline
& & &\\
$t_{\kappa}\ll t\ll t_a$ & $\displaystyle \sqrt{\frac{t_{\kappa}}{\pi}}\frac{
N_0}{t^{1/2}}$ & $\displaystyle 2D_st_{\kappa}+\frac{2\sqrt{t_{\kappa}}}{
\sqrt{\pi}}D_bt^{1/2}$ & $\displaystyle 2\sqrt{\pi t_{\kappa}}D_st^{1/2}+2
D_bt$ \\
& & &\\
\hline
& & &\\
$t_a\ll t\ll t_b$ & $\displaystyle \frac{1}{2}\sqrt{t_at_{\kappa}}\frac{N_0}{t}$
& $\displaystyle t_at_{\kappa}D_s\frac{1}{t}\ln\left(\frac{4t}{Ct_a}\right)+
\sqrt{t_at_{\kappa}}D_b$ & $\displaystyle 2\sqrt{t_at_{\kappa}}D_s\ln\left(
\frac{4t}{Ct_a}\right)+2D_bt$\\
& & &\\
\hline
& & &\\
$t_b\ll t$ & $\displaystyle2\frac{\sqrt{t_at_{\kappa}}}{t_b}N_0$ &
$\displaystyle \frac{8t_at_{\kappa}}{t_b^2}D_st+\frac{4\sqrt{t_at_{\kappa}}}{
t_b}D_bt$ & $\displaystyle 4\frac{\sqrt{t_at_{\kappa}}}{t_b}D_st+2D_bt$\\
& & &\\
\hline
\end{tabular}
\caption{Effective surface diffusion, strong coupling limit. For the
different regimes we list the number of particles $N_s(t)$ on the surface,
the surface mean squared displacement $\langle z^2(t)\rangle$, and the
normalized mean squared displacement $\langle z^2(t)\rangle_{\mathrm{norm}}$.
$C=\exp(\gamma)\approx1.78107$ where $\gamma$ is Euler's constant.}
\label{tab_strong}
\end{table*}

The mean squared displacement has a very interesting initial anomalously
diffusive behavior $\simeq t^{3/2}$ \cite{report,igor}. This superdiffusion
arises due to
mediation by bulk excursions resulting in the effective Cauchy distribution
\begin{equation}
n(z,t)\sim\frac{N_0\kappa t}{\pi\left(z^2+\kappa^2 t^2\right)}.
\end{equation}
In this initial regime we can use a simple scaling argument to explain this
superdiffusive behavior, compare the discussion in Ref.~\cite{bychuk}. Thus,
once detached from the surface a particle returns to the surface with a
probability distributed according to $\simeq t^{-1/2}$. Due to the diffusive
coupling $z^2\simeq t$ in the bulk the effective displacement along the
cylinder is then distributed according to $\simeq |z|^{-1}$, giving rise to
a probability density $\simeq z^{-2}$.

Later, the mean squared displacement turns over to a square root behavior
corresponding to subdiffusion. As can be seen from the associated normalized
mean squared displacement, this behavior is due to the escaping particles.
At even longer times the mean squared displacement reaches a plateau value.
This is a remarkable property of this cylindrical geometry, reflecting a
delicate balance between decreasing particle number and increasing length of
the bulk mediated surface translocations. This plateau is the \emph{terminal\/}
behavior when no outer cylinder is present. That is, even at infinite times,
when fewer and fewer particles are on the surface, the surface mean squared
displacement does not change. In presence of the outer cylinder the mean
squared displacement eventually is dominated by the bulk motion and acquires
the normal linear growth with time.

Combining the dynamics of the number of surface particles and the mean
squared displacement we obtain the behavior of the normalized mean squared
displacement listed in the last column.

\subsubsection{Intermediate coupling limit}

In the intermediate coupling limit the results are listed in Table
\ref{tab_int}. Also in this regime we observe the initial superdiffusion and
associated Cauchy form of the surface particle concentration. The subsequent
regime of intermediate times splits up into two subregimes. This subtle
turnover will be discussed in detail below. The last two regimes exhibit the
same behavior as the corresponding regimes in the strong coupling limit.

\begin{table*}
\begin{tabular}{|c|c|c|c|}
\hline
& & &\\
\hspace*{0.6cm} Time regime \hspace*{0.6cm} & \hspace*{0.8cm} $N_s(t)$
\hspace*{0.8cm} & \hspace*{2.0cm} $\langle z^2(t)\rangle$  \hspace*{2.0cm} &
\hspace*{1.8cm} $\langle z^2(t)\rangle_{\mathrm{norm}}$ \hspace*{1.8cm}
\\
& & &\\
\hline
& & &\\
$t\ll t_a$ & $N_0$ & $\displaystyle2D_st+\frac{4}{3\sqrt{\pi t_{\kappa}}}
D_bt^{3/2}$ &
$\displaystyle2D_st+\frac{4}{3\sqrt{\pi t_{\kappa}}}D_bt^{3/2}$\\
& & &\\
\hline
& & &\\
$t_a\ll t<t_c\ll t_{\kappa}$ & $N_0$ &
$\displaystyle2D_st+\frac{2D_bt^2}{t_c\ln^2\left(4t/[C^2t_a]\right)}$ &
$\displaystyle2D_st+\frac{2D_bt^2}{t_c\ln^2\left(4t/[C^2t_a]\right)}$\\
& & &\\
$t_a\ll t_c<t\ll t_{\kappa}$ & transition to $\simeq1/t$ &
transition to $\simeq\mathrm{const}$ & transition to $\simeq t$\\
& & &\\
\hline
& & &\\
$t_{\kappa}\ll t\ll t_b$ & $\displaystyle\frac{1}{2}\sqrt{t_at_{\kappa}}
\frac{N_0}{t}$ &
$\displaystyle t_at_{\kappa}D_s\frac{1}{t}\ln\left(\frac{4t}{Ct_a}\right)
+\sqrt{t_at_{\kappa}}D_b$ &
$\displaystyle2\sqrt{t_at_{\kappa}}D_s\ln\left(\frac{4t}{Ct_a}\right)+2D_bt$\\
& & &\\
\hline
& & &\\
$t_b\ll t$ &
$\displaystyle\frac{2\sqrt{t_at_{\kappa}}}{t_b}N_0$ &
$\displaystyle\frac{8t_at_{\kappa}}{t_b^2}D_st+4\frac{\sqrt{t_at_{\kappa}}}{
t_b}D_bt$ &
$\displaystyle4\frac{\sqrt{t_at_{\kappa}}}{t_b}D_st+2D_bt$\\
& & &\\
\hline
\end{tabular}
\caption{Effective surface diffusion, intermediate coupling limit.}
\label{tab_int}
\end{table*}

\subsection{Numerical evaluation}

\begin{figure}
\includegraphics[width=8.8cm]{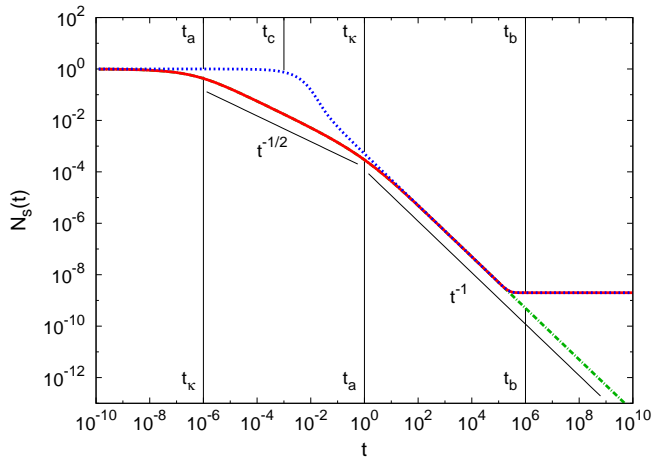}
\caption{Time evolution of the number of surface particles obtained by numerical
Laplace inversion, for the following cases: strong binding with (green
dashed-dotted line) and without (red full line) outer cylinder, and
intermediate binding in presence of the outer cylinder (blue dashed line).
The indicated characteristic time scales correspond to the cases of strong
(bottom) and intermediate (top) coupling.}
\label{num}
\end{figure}

\begin{figure*}
\includegraphics[width=8.8cm]{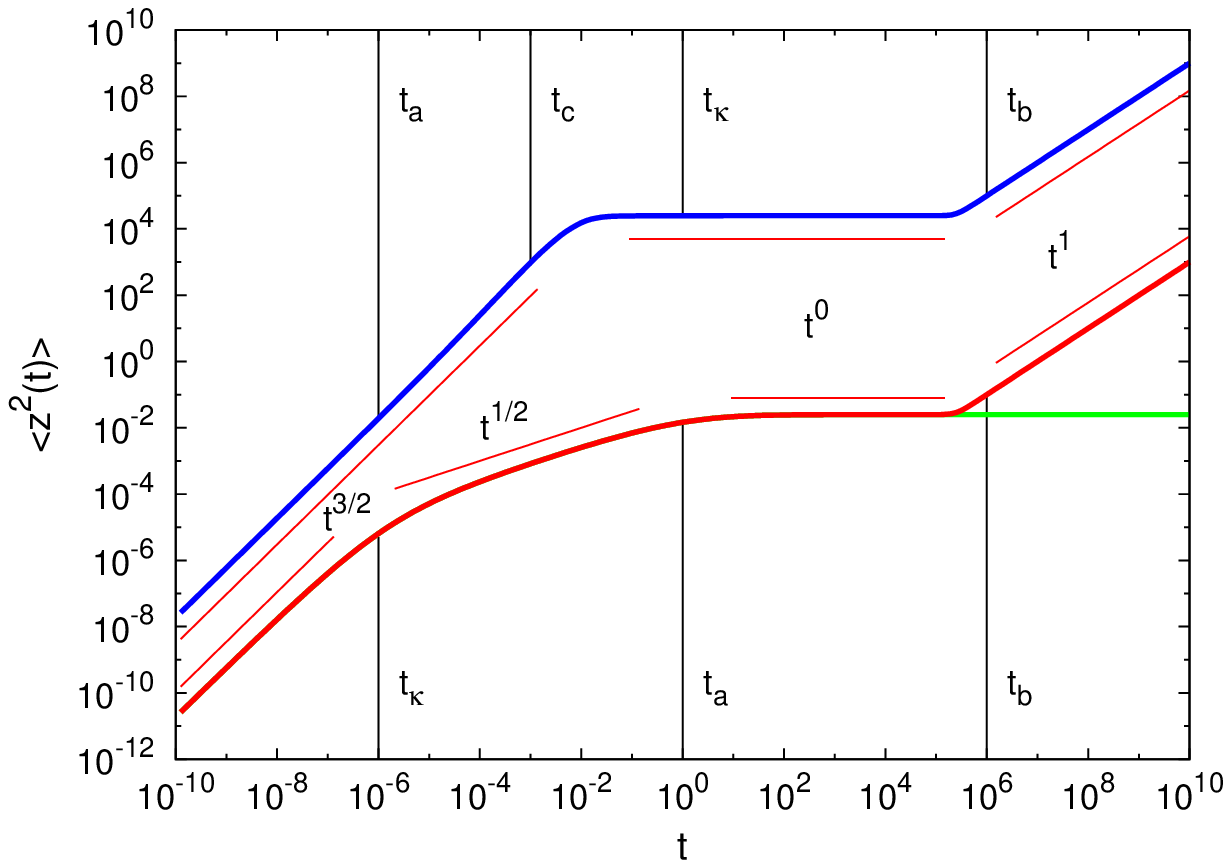}
\includegraphics[width=8.8cm]{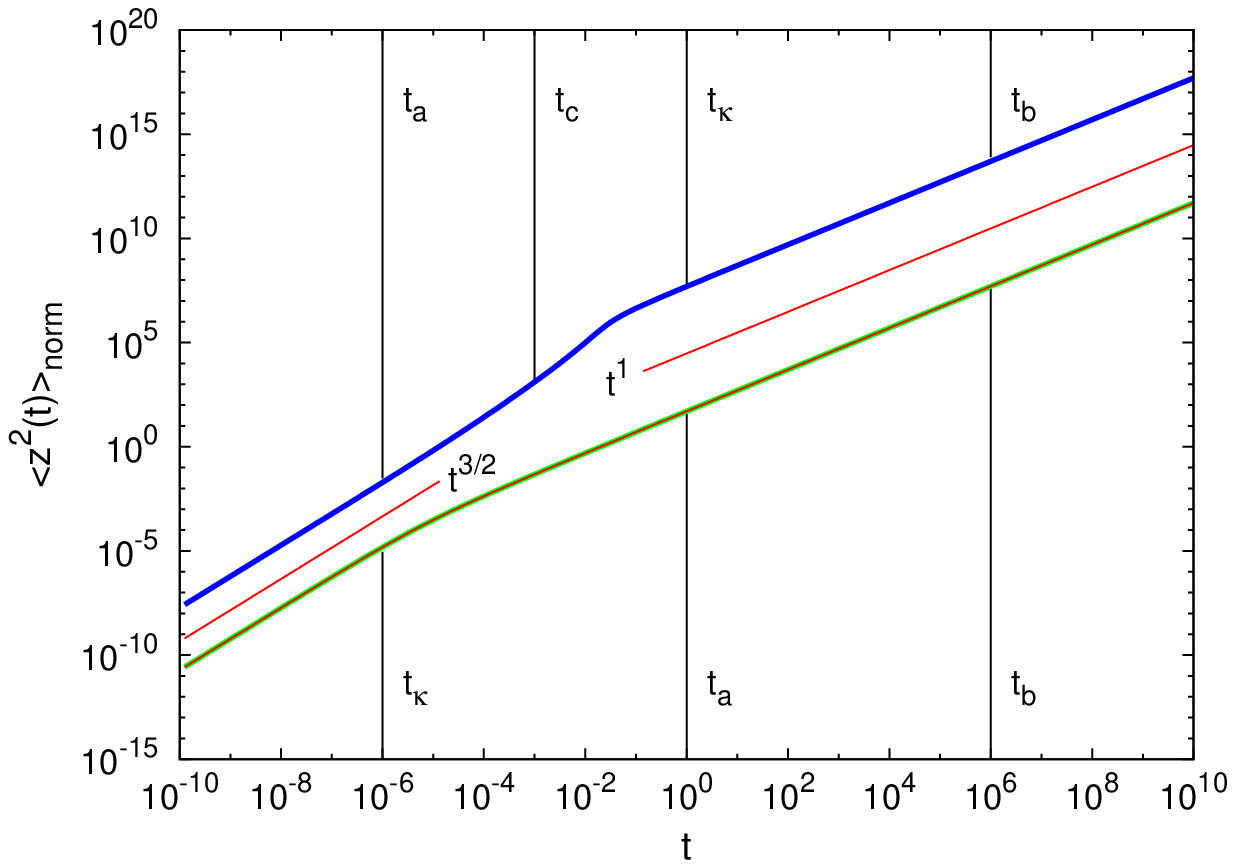}
\caption{Surface mean squared displacement obtained by numerical Laplace
inversion. Left: The graphs show the various effective diffusion regimes
$\langle z^2(t)\rangle$ along the cylinder. Note the characteristic
transient plateau, which is the terminal behavior in the absence of a
confining outer cylinder. Right: Normalized function $\langle z^2(t)\rangle_
{\mathrm{norm}}$. The subregime $t_c<t<t_a$ in the intermediate case
distinctly shows a superdiffusive behavior that is even steeper than the
initial $t^{3/2}$ scaling. We show the strong binding case with (red line)
and without (green line) and outer cylinder, as well as the case of
intermediate binding with outer cylinder (blue line). The characteristic
time scales connected with the associated curves denote the cases of strong
(bottom) and intermediate (top) coupling.}
\label{msd}
\end{figure*}

In Figs.~\ref{num} and \ref{msd} we show results from numerical Laplace
inversion of the exact expressions for the number of surface particles
and the surface mean squared displacement. We consider both the strong
and intermediate coupling cases. The parameters fixing the time scales
were chosen far apart
from each other to distinguish the different limiting behaviors
computed in the following Sections. In all figures a vanishing surface
diffusivity ($D_s=0$) is chosen for clarity.

For strong coupling the selected time scales are $t_{\kappa}=10^{-6}$, $t_a
=1$, and $t_b=10^6$ in dimensionless units. Therefore the bulk diffusion
constant becomes $D_b=a^2/t_a=25$ for our choice $a=5$. The coupling constant
is $\kappa=a/\sqrt{t_at_{\kappa}}=5\times10^3$, and the outer cylinder radius
becomes $b=a/\sqrt{t_b/t_a}=5\times10^3$.

In the intermediate regime we chose $t_a=10^{-6}$, $t_{\kappa}=1$, and $t_b=
10^6$. This sets the bulk diffusivity to $D_b=25\times10^6$ and the outer
cylinder radius to $b=5\times10^6$. These values are chosen such that we
can plot the results for the intermediate case alongside the strong coupling
case.

Fig.~\ref{num} shows the time evolution of the number of surface particles,
normalized to $N_0=1$. For the strong coupling case the value remains almost
constant until $t\approx t_{\kappa}$, and then turns over to an inverse square
root decay that lasts until $t\approx t_a$. Subsequently a $t^{-1}$ behavior
emerges. In presence of an outer cylinder, due to the confinement this
inversely time proportional evolution is finally terminated by a stationary
plateau. In the intermediate coupling case similar behavior is observed,
apart from the two subregimes in the range at intermediate times.

Fig.~\ref{msd} depicts the behavior of the surface mean squared displacement.
In the left panel the function $\langle z^2(t)\rangle$ shows the various
regimes found in the strong and intermediate coupling limits. Remarkably
the intermediate coupling regime exhibits a superdiffusive behavior in the
range $t_c<t<t_a$ that is even faster than the initial $t^{3/2}$ scaling.
The right panel of Fig.~\ref{msd} shows the behavior of the normalized surface
mean squared displacement. See Sections \ref{sec_scl} and \ref{sec_icl} for
details.

\section{Coupled diffusion equations and general solution}
\label{results}

In this Section we state the polar symmetry of the problem we want to
consider, and then formulate the starting equations for our model. The
general solution is presented in Fourier-Laplace space. In the two
subsequent Sections we calculate explicit results in various limiting
cases, for strong and intermediate coupling.

\subsection{Starting equation and particle number conservation}

The full problem is spanned by the coordinates $z$ measured along the
cylinder axis, the radius $r$ measured perpendicular to the $z$ axis,
and the corresponding polar angle $\theta$. We are only interested in
the effective displacement of particles along the cylinder axis and
therefore eliminate the $\theta$ dependence. This can be consistently
done in the following way. (i) As initial condition we assume that initially
the particles are concentrated as a sharp $\delta(z)$ peak on the inner
cylinder surface, homogeneous in the angle coordinate $\theta$.
(ii) Our boundary conditions are $\theta$ independent.

For the bulk concentration of particles in the volume between the inner and
outer cylinders this symmetry requirement simply means that we can integrate
out the $\theta$ dependence and consider this concentration as function of
$z$, radius $r$, and time $t$: $\mathcal{C}=\mathcal{C}(r,z,t)$. The physical
dimension of the concentration $\mathcal{C}$ is $[\mathcal{C}]=1/\mathrm{cm}
^3$. On the surface of the inner cylinder we measure the concentration by the
density $n_{\mathrm{2D}}(z,t)$, which is of dimension $[n_{\mathrm{2D}}]=1/
\mathrm{cm}^2$. Note that $n_{\mathrm{2D}}(z,t)$ does not explicitly depend
on $\theta$. We average this cylinder surface density over the polar angle,
and obtain the line density $n(z,t)$:
\begin{equation}
n(z,t)=a\int_0^{2\pi}n_{\mathrm{2D}}(z,t)d\theta=2\pi an_{\mathrm{2D}}(z,t),
\end{equation}
such that $[n]=1/\mathrm{cm}$. Note that on the inner cylinder with radius
$a$ the expression $ad\theta dz$ is the cylindrical surface increment. The
factor $2\pi a$ is important when we formulate the reactive boundary
condition on the inner cylinder connecting surface line density $n(z,t)$
and the volume density $\mathcal{C}(r,z,t)$.

Given the line density $n$, the total number $N_s(t)$ of particles on the
inner cylinder surface at given time $t$ becomes
\begin{equation}
N_s(t)=\int_0^{2\pi}ad\theta\int_{-\infty}^{\infty}n_{\mathrm{2D}}(z,t)dz=
\int_{-\infty}^{\infty}n(z,t)dz.
\end{equation}
We assume that initially $N_0$ particles are concentrated in a $\delta$-peak
on the cylinder surface at $z=0$:
\begin{equation}
\label{ninit}
n(z,t)\Big|_{t=0}=N_0\delta(z).
\end{equation}
Consequently the initial bulk concentration vanishes everywhere on the
interval $a<r\le b$ such that
\begin{equation}
\label{cinit}
\mathcal{C}(r,z,t)\Big|_{t=0}=0.
\end{equation}

Let us now specify the boundary conditions at the two cylinder surfaces.
At the outer cylinder ($r=b$) we impose a reflecting boundary condition
of the Neumann form
\begin{equation}
\label{reflective}
\frac{\partial}{\partial r}\mathcal{C}(r,z,t)\Big|_{r=b}=0.
\end{equation}
In the case when we do not consider an outer cylinder ($b\to\infty$) this
Neumann condition may be replaced by a natural boundary condition of the
form
\begin{equation}
\lim_{r\to\infty}\mathcal{C}(r,z,t)=0.
\end{equation}
The reactive boundary condition on the inner cylinder ($r=a$) is derived
from a discrete random walk process in App.~\ref{reactive_bc} (compare
also Refs.~\cite{subsurf}). Accordingly we balance the flux away from the
inner cylinder surface,
\begin{equation}
j_{\mathrm{off}}=\frac{1}{\tau_{\mathrm{off}}}n_{\mathrm{2D}}(z,t)=\frac{1}{
2\pi a\tau_{\mathrm{off}}}n(z,t),
\end{equation}
by the incoming flux from the bulk onto the cylinder surface,
\begin{equation}
\label{rbc1}
j_{\mathrm{on}}=\lim_{r\to a}k_b\mathcal{C}(r,z,t).
\end{equation}
Here, $\tau_{\mathrm{off}}$ with dimension $[\tau_{\mathrm{off}}]=1/\mathrm{
sec}$ is the characteristic time scale for particle unbinding from the surface.
It is proportional to the Arrhenius factor of the binding free energy
$\varepsilon$ of the particles, $\exp(-|\varepsilon|/[k_BT])$, where $k_BT$
denotes the thermal energy at temperature $T$. The binding rate $k_b$, in
contrast,
has physical dimension $[k_b]=\mathrm{cm}/\mathrm{sec}$, which is typical for
surface-bulk coupling in cylindrical coordinates, compare the discussions in
Refs.~\cite{bvh,berg,subsurf}. For convenience, we collect the coefficients
in the reactive boundary condition (\ref{rbc1}) into the coupling constant
\begin{equation}
\mu\equiv\frac{1}{2\pi ak_b\tau_{\mathrm{off}}},
\end{equation}
such that our reactive boundary condition finally is recast into the form
\begin{equation}
\label{reactive}
\mathcal{C}(r,z,t)\Big|_{r=a}=\mu n(z,t).
\end{equation}

The time evolution of the bulk density $\mathcal{C}(r,z,t)$ is governed by
the cylindrical diffusion equation
\begin{equation}
\label{ceq}
\frac{\partial}{\partial t}\mathcal{C}(r,z,t)=D_b\left(\frac{1}{r}\frac{
\partial}{
\partial r}\left[r\frac{\partial}{\partial r}\right]+\frac{\partial^2}{
\partial z^2}\right)\mathcal{C}(r,z,t),
\end{equation}
valid on the domain $a<r<b$ and $-\infty<z<\infty$. In Eq.~(\ref{ceq}), $D_b$
is the bulk diffusion coefficient of dimension $[D_b]=\mathrm{cm^2}/\mathrm{
sec}$. From a random walk perspective we can write $D_b=\langle\delta\xi^2
\rangle/(6\langle\delta\tau\rangle)$, where $\langle\delta\xi^2\rangle$ is the
average variance of individual jumps, and $\langle\delta\tau\rangle$ is the
typical time between consecutive jumps. As shown in App.~\ref{reactive_bc}
the dynamic equation for the line density $n$ directly includes the incoming
flux term and is given by
\begin{equation}
\label{ndiffeq}
\frac{\partial}{\partial t}n=D_s\frac{\partial^2}{\partial z^2}n(z,t)
+2\pi aD_b\frac{\partial}{\partial r}\mathcal{C}(r,z,t)\Big|_{r=a},
\end{equation}
where $D_s$ denotes the surface diffusion coefficient. In many realistic cases
the magnitude of $D_s$ is considerably smaller than the bulk diffusivity
$D_b$. The coupling term connects the surface density $n$ to the bulk
concentration $\mathcal{C}$. The fact that here the bulk diffusivity
occurs as coupling term stems from the continuum limit, in which the
binding rate diverges, and therefore the binding corresponds to the
step from the exchange site to the surface.

The diffusion equations (\ref{ceq}) and (\ref{ndiffeq}) together with the
boundary conditions (\ref{reflective}) and (\ref{reactive}) as well as
the initial conditions (\ref{ninit}) and (\ref{cinit}) completely specify
our problem. Moreover the total number of particles is conserved. Namely,
the number of surface particles varies with time as
\begin{equation}
\frac{dN_s(t)}{dt}=2\pi aD_b\int_{-\infty}^{\infty}\frac{\partial}{\partial r}
\mathcal{C}(r,z,t)\Big|_{r=a}dz,
\end{equation}
as can be seen from integration of Eq.~(\ref{ndiffeq}) over $z$ and noting
that $n(|z|\to\infty,t)=0$. For the number of bulk particles we obtain
\begin{widetext}
\begin{eqnarray}
\nonumber
\frac{dN_b(t)}{dt}&=&2\pi D_b\int_a^brdr\int_{-\infty}^{\infty}dz\,\frac{1}{r}
\frac{\partial}{\partial r}\left(r\frac{\partial}{\partial r}\mathcal{C}(r,
z,t)\right)
=2\pi D_b\int_{-\infty}^{\infty}\left[\left(r\frac{\partial}{\partial r}
\mathcal{C}(r,z,t)\right)_{r=b}-\left(r\frac{\partial}{\partial r}\mathcal{C}(
r,z,t)\right)_{r=a}\right]\\
&=&-2\pi aD_b\int_{-\infty}^{\infty}\frac{\partial}{\partial r}
\mathcal{C}(r,z,t)\Big|_{r=a}dz.
\end{eqnarray}
\end{widetext}
From these two relations we see that indeed the total number of particles
fulfills
\begin{equation}
\frac{d}{dt}\Big(N_s(t)+N_b(t)\Big)=0,
\end{equation}
and therefore $N_s(t)+N_b(t)=N_0$.

\subsection{Solution of the bulk diffusion equation}

To solve Eq.~(\ref{ceq}) and the corresponding boundary and initial value
problem we use the Fourier-Laplace transform method.
The dynamic equation for $\mathcal{C}(r,k,s)$ is the
ordinary differential equation
\begin{equation}
\label{ceq_fl}
\frac{d^2}{dr^2}\mathcal{C}(r,k,s)+\frac{1}{r}\frac{d}{dr}\mathcal{C}(r,k,s)
-q^2\mathcal{C}(r,k,s)=0,
\end{equation}
where we use the abbreviation
\begin{equation}
q^2=k^2+\frac{s}{D_b}.
\end{equation}
The reactive boundary condition becomes
\begin{equation}
\mathcal{C}(r,k,s)\Big|_{r=a}=\mu n(k,s),
\end{equation}
and for the reflective condition we find
\begin{equation}
\frac{d}{dr}\mathcal{C}(r,k,s)\Big|_{r=b}=0.
\end{equation}

The general solution of Eq.~(\ref{ceq_fl}) is given in terms of the zeroth
order modified Bessel functions $I_0$ and $K_0$ in the linear combination
\begin{equation}
\mathcal{C}(r,k,s)=AI_0(qr)+BK_0(qr).
\end{equation}
The constants $A$ and $B$ follow from the boundary conditions, such that
\begin{equation}
AI_0(qa)+BK_0(qa)=\mu n(k,s)
\end{equation}
and
\begin{equation}
A\frac{\partial}{\partial r}I_0(qr)\Big|_{r=b}+B\frac{\partial}{\partial r}
K_0(qr)\Big|_{r=b}=0.
\end{equation}
Using $\partial I_0(qr)/\partial r=qI_1(qr)$ and $\partial K_0(qr)/\partial
r=-qK_1(qr)$, we can rewrite the latter relation:
\begin{equation}
AqI_1(qb)-BqK_1(qb)=0.
\end{equation}
The two coefficients are therefore given by
\begin{equation}
A=\frac{K_1(qb)}{\Delta(k,s)}\mu n(k,s),\quad B=\frac{I_q(qb)}{\Delta(k,s)}
\mu n(k,s),
\end{equation}
where we introduce the abbreviation
\begin{equation}
\label{delta_abb}
\Delta(k,s)\equiv I_0(qa)K_1(qb)+I_1(qb)K_0(qa).
\end{equation}
Note that, due to the definition of the variable $q$ the function $\Delta$
indeed explicitly depends on the Laplace variable $s$. The solution for the
bulk density $\mathcal{C}$ in Fourier-Laplace domain is therefore given by
the expression
\begin{equation}
\label{csol_fl}
\mathcal{C}(r,k,s)=\frac{\mu n(k,s)}{\Delta(k,s)}\Big(K_1(qb)I_0(qr)+
I_1(qb)K_0(qr)\Big).
\end{equation}

\subsection{Solution of the surface diffusion equation}

In a similar fashion we obtain the Fourier-Laplace transform of the dynamic
equation for the surface density $n$, namely
\begin{equation}
sn(k,s)-N_0=-k^2D_sn(k,s)+2\pi aD_b\frac{\partial}{\partial r}\mathcal{C}(
r,k,s)\Big|_{r=a}.
\end{equation}
Defining the propagator of the homogeneous equation,
\begin{equation}
G_s(k,s)=\frac{1}{s+k^2D_s},
\end{equation}
we find
\begin{equation}
\label{n_prop}
n(k,s)=N_0G_s(k,s)+G_s(k,s)2\pi aD_b\frac{\partial}{\partial r}\mathcal{C}(
r,k,s)\Big|_{r=a}.
\end{equation}
From Eq.~(\ref{csol_fl}) we obtain for the reactive boundary condition that
\begin{equation}
\label{rbc}
\frac{\partial}{\partial r}\mathcal{C}(r,k,s)\Big|_{r=a}=-\mu n(k,s)\frac{q
\Delta_1(k,s)}{\Delta(k,s)},
\end{equation}
where
\begin{equation}
\label{delta1_abb}
\Delta_1(k,s)\equiv K_1(qa)I_1(qb)-I_1(qa)K_1(qb).
\end{equation}
Insertion of relation (\ref{rbc}) into Eq.~(\ref{n_prop}) produces the
result
\begin{equation}
\label{nsol_fl}
n(k,s)=\frac{N_0}{\displaystyle s+k^2D_s+\kappa q\frac{\Delta_1(k,s)}{
\Delta(k,s)}}.
\end{equation}
Here we also define the coupling constant
\begin{equation}
\label{kappa}
\kappa\equiv2\pi a\mu D_b=\frac{D_b}{k_b\tau_{\mathrm{off}}},
\end{equation}
which allows us to distinguish the regimes of strong, intermediate, and
weak bulk-surface coupling used in this work. If we remove the outer
cylinder, that enforces a finite cross-section in the cylindrical symmetry,
we obtain the following simplified expression,
\begin{equation}
n(k,s)=\frac{N_0}{\displaystyle s+k^2D_s+\kappa q\frac{K_1(qa)}{K_0(qa)}},
\end{equation}
as in the limit $b\to\infty$, we have $I_{\nu}(qb)\to\infty$ and $K_{\nu}
(qb)\to0$. From the Fourier-Laplace transform (\ref{nsol_fl}) the number
of particles on the cylinder surface is given by
\begin{equation}
N_s(s)=n(k=0,s),
\end{equation}
following the definition of the Fourier transform.

Plugging the result (\ref{nsol_fl}) into Eq.~(\ref{csol_fl}) we obtain the
closed form for the Fourier-Laplace transform of the bulk concentration,
\begin{widetext}
\begin{eqnarray}
\nonumber
\mathcal{C}(r,k,s)&=&\frac{\mu N_0}{\Delta(k,s)\Big(s+k^2D_s+\kappa q\Delta_1(
k,s)/\Delta(k,s)\Big)}\Big(K_1(qb)I_0(qr)+I_1(qb)K_0(qr)\Big)\\
&=&\frac{N_0}{2\pi ak_b\tau_{\mathrm{off}}}\frac{K_1(qb)I_0(qr)+I_1(qb)K_0(qr)}
{\Delta(k,s)\Big(s+k^2D_s+\kappa q\Delta_1(k,s)/\Delta(k,s)\Big)}.
\end{eqnarray}

We note that the solutions for $n(k,s)$ and $\mathcal{C}(k,s)$ indeed fulfill
the particle conservation,
\begin{equation}
\int_{-\infty}^{\infty}n(z,t)dz+2\pi\int_{-\infty}^{\infty}dz\int_a^brdr\,
\mathcal{C}(r,z,t)=N_0 \,\,\,\Leftrightarrow\,\,\,
n(k,s)\Big|_{k=0}+2\pi\int_a^brdr\,\mathcal{C}(r,k,s)\Big|_{k=0}=\frac{N_0}{s}.
\end{equation}
\end{widetext}

Using the results for the surface propagator $n(z,t)$, Eq.~(\ref{nsol_fl}),
we characterize the effective surface diffusion on the cylinder in terms of
the single-particle mean squared displacement
\begin{equation}
\label{surfmsd}
\langle z^2(t)\rangle=N_0^{-1}\int_{-\infty}^{\infty}z^2n(z,t)dz.
\end{equation}
In Fourier-Laplace domain, we re-express this integral as
\begin{equation}
\label{msd_f}
\langle z^2(s)\rangle=-N_0^{-1}\left.\frac{\partial^2n(k,s)}{\partial k^2}
\right|_{k=0}.
\end{equation}
This mean squared displacement includes the unbinding dynamics of particles
as manifest in the quantity $N_s(t)$. We can exclude this effect by defining
the normalized mean squared displacement
\begin{equation}
\langle z^2(t)\rangle_{\mathrm{norm}}=\frac{N_0}{N_s(t)}\langle z^2(t)\rangle.
\end{equation}

From above results for the effective surface propagator we obtain the exact
result for the surface mean squared displacement in App.~\ref{msd_calc}. In
what follows, however, for simplicity of the argument we proceed differently.
Namely we first approximate the effective surface propagator $n(z,t)$, and
from the various limiting forms determine the surface mean squared
displacement. Comparison to the limits taken from the general results derived
in App.~\ref{msd_calc}, it can be shown that both procedures yield identical
results.

\section{Explicit calculations: strong coupling limit}
\label{sec_scl}

In this Section we consider the strong coupling limit $t_{\kappa}\ll t_a\ll
t_b$, representing the richest of the three regimes. Based on the result
for the effective surface propagator, Eq.~(\ref{nsol_fl}), in Fourier-Laplace
space obtained in the previous Section we now calculate the quantities
characteristic of the effective motion on the cylinder surface, as mediated
by transient bulk excursions. We consider the number of particles
on the surface, the axial mean squared displacement, as well as the surface
propagator. We divide the discussion into the four different dynamic regimes
defined by comparison of the involved time scales $t_{\kappa}$, $t_a$, and
$t_b$.

\subsection{Short times, $t\ll t_{\kappa}\ll t_a\ll t_b$}
\label{scl_short}

The short time limit $t\ll t_{\kappa}\ll t_a\ll t_b$ corresponds to the
Laplace domain regime
\begin{equation}
st_{\kappa},st_a,st_b\gg1.
\end{equation}

\subsubsection{Surface propagator in Fourier-Laplace space}

We first obtain the short time limit of the effective surface propagator in
Fourier-Laplace space. To this end we note that the following inequalities
hold:
\begin{equation}
qa=a\sqrt{k^2+\frac{s}{D_b}}\ge a\sqrt{\frac{s}{D_b}}=\sqrt{st_a}\gg1,
\end{equation}
and thus we have
\begin{equation}
qa\gg1\,\,\,\mbox{and}\,\,\,qb\gg1.
\end{equation}
For this case we use the following expansion of the Bessel functions contained
in the abbreviations $\Delta(k,s)$ and $\Delta_1(k,s)$. Namely, for $z\to
\infty$,
\begin{equation}
I_{\nu}(z)\sim\frac{\exp(z)}{\sqrt{2\pi z}},\,\,\,K_{\nu}(z)\sim\sqrt{\frac{
\pi}{2z}}\exp(-z).
\label{bessel_moreexp}
\end{equation}
From expressions (\ref{delta_abb}) and (\ref{delta1_abb}) we find
\begin{equation}
\Delta(k,s)\sim\Delta_1(k,s)\sim\frac{\exp(q[b-a])}{2q\sqrt{ab}}.
\end{equation}
Therefore, the surface propagator in Fourier-Laplace in the short time limit
reduces to the simplified form
\begin{equation}
\label{limprop}
n(k,s)\sim\frac{N_0}{s+k^2D_s+\kappa\sqrt{k^2+s/D_b}}.
\end{equation}

\subsubsection{Number of particles on the surface}

From the relation $N_s(s)=n(k=0,s)$ we obtain the number of surface
particles by help of the above expression for the limiting form of $n(k,s)$:
\begin{equation}
\label{nsurf_1}
N_s(s)\sim\frac{N_0t_{\kappa}}{st_{\kappa}+\sqrt{st_{\kappa}}}.
\end{equation}
Since $st_{\kappa}\gg1$ the leading behavior follows
\begin{equation}
\label{N_lead}
N_s(s)\sim\frac{N_0}{s},
\end{equation}
i.e., we recover that the number of particles on the surface remains
approximately conserved in the short time regime,
\begin{equation}
N_s(t)\sim N_0.
\end{equation}

\subsubsection{Surface mean squared displacement}

The surface mean squared displacement is readily obtained from the limiting
form of the surface propagator (\ref{limprop}) by help of relation
(\ref{msd_f}). Namely, we obtain
\begin{equation}
\label{surf_scl_1}
\langle z^2(s)\rangle\sim\frac{2D_s+D_b/\sqrt{st_{\kappa}}}{\left(s+s/\sqrt{
st_{\kappa}}\right)^2}.
\end{equation}
Since $st_{\kappa}\gg1$ the leading behavior corresponds to
\begin{equation}
\langle z^2(s)\rangle\sim\frac{2D_s}{s^2}+\frac{D_b}{t_{\kappa}^{1/2}s^{5/2}},
\end{equation}
from which the time-dependence
\begin{equation}
\label{msd_res1}
\langle z^2(t)\rangle\sim\langle z^2(t)\rangle_{\mathrm{norm}}\sim2D_st\left[1+\frac{2}{3\pi^{1/2}}\frac{D_b}{D_s}\left(\frac{t}{t_{\kappa}}\right)^{1/2}
\right].
\end{equation}
yields after Laplace inversion. As in this short time regime $N_s(t)\sim N_0$,
the normalized surface mean squared displacement follows the same behavior.

Remarkably, result (\ref{msd_res1}) contains a contribution growing like
$\simeq t^{3/2}$. This superdiffusive behavior becomes relevant when $D_b
\gg D_s$, which is typically observed in many systems. Thus for DNA binding
proteins the bulk diffusivity may be a factor of $10^2$ or more larger than the
diffusion constant along the DNA: for Lac repressor the bulk diffusivity
is of the order of $5..9\times10^{-7}\mathrm{cm}^2/\mathrm{sec}$, while the
one-dimensional diffusion constant along the DNA surface ranges in between
$2..9\times10^{-10}\mathrm{cm}^2/\mathrm{sec}$ \cite{wang,winter}.

\subsubsection{Surface propagator in real space}

We now turn to the functional form of the surface propagator $n(z,t)$ in
real space at short times in the strong coupling limit. We investigate
this quantity in the limit $D_s=0$ of vanishing surface diffusion.

In the current short time limit $t\ll t_{\kappa}\ll t_a\ll t_b$ we distinguish
two parts of the surface density $n$. Let us start with the central part defined
by $k^2\gg s/D_b$. The corresponding limiting form of Eq.~(\ref{limprop}) is
then given by
\begin{equation}
n(k,s)\sim\frac{N_0}{s+\kappa|k|}.
\end{equation}
The inverse Fourier-Laplace transform leads to the Cauchy probability
density function
\begin{equation}
\label{cauchy}
n(z,t)\sim\frac{N_0\kappa t}{\pi\left(z^2+\kappa^2t^2\right)}.
\end{equation}
This central part of the surface propagator obeys the governing dynamic
equation \cite{report,chechkin_jsp}
\begin{equation}
\label{sfde}
\frac{\partial}{\partial\kappa t}n(z,t)=\frac{\partial}{\partial|z|}n(z,t)
\end{equation}
with initial condition $n(z,t=0)=N_0\delta(z)$. Here, we defined the space
fractional derivative $\partial/\partial|z|$ in the Riesz-Weyl sense whose
Fourier transform takes on the simple form \cite{samko}
\begin{equation}
\int_{-\infty}^{\infty}e^{ikz}\left(\frac{\partial}{\partial|z|}n(z,t)
\right)dz=-|k|n(k,t).
\end{equation}

Eq.~(\ref{cauchy}) and the corresponding dynamic equation (\ref{sfde}) are
remarkable results, which are analogous to the findings in Ref.~\cite{bychuk1}
for a flat surface obtained from scaling arguments \cite{REMM}. It says
that the bulk mediation causes an effective surface motion whose propagator
is a L{\'e}vy stable law of index 1. This behavior can be guessed from the
scaling of the returning probability to the surface, together with the
diffusive scaling $z^2\simeq t$. However, the resulting Cauchy distribution
cannot have an infinite range, as the particle in a finite time only
diffuses a finite distance. The question therefore arises whether there
exists a cutoff of the Cauchy law, and of what form this is.

The advantage of our exact treatment is that the Cauchy law can be derived
explicitly, but especially the transition to other regimes studied. To this
end we introduce the time-dependent length scale
\begin{equation}
\ell_C(t)=\sqrt{D_bt}
\end{equation}
which turns out to define the range of validity of the Cauchy region. Namely,
while at distances $z>\ell_C(t)$ we observe a cutoff of the Cauchy behavior,
for $z<\ell_C(t)$ the Cauchy approximation is valid.
Note that in this short time regime $t<t_{\kappa}$ the Cauchy range scales
as $z_{\mathrm{max}}\approx\sqrt{D_bt}$ such that $z^2$ can indeed become
significantly larger than
$\kappa^2t^2$ at sufficiently short times, and thus the power law asymptotics
$n(z,t)\simeq z^{-2}$ in Eq.~(\ref{cauchy}) become relevant.
From this Cauchy part we obtain the superdiffusive contribution
\begin{equation}
\int_{-\ell_C(t)}^{\ell_C(t)}\frac{z^2\kappa t\,dz}{\pi\left(z^2+\kappa^2t^2
\right)}\approx\frac{2}{\pi}\kappa\sqrt{D_b}t^{3/2}
\end{equation}
to the mean squared displacement, that is consistent with the exact forms
(\ref{msd_res1}) and
(\ref{shortmsd}) [with $D_s=0$]. Calculation of the mean squared displacement
however requires the $k\to0$ limit and thus involves the extreme wings of the
distribution. As the system evolves in time the central Cauchy part spreads.
Already in the regime $t_{\kappa}<t<t_a$ we have $D_bt<\kappa^2t^2$, and the
asymptotic behavior $\simeq z^{-2}$ can no longer be observed.

To show how at very large $|z|$ the Cauchy form of the propagator is truncated
we consider Eq.~(\ref{limprop}) for small wave number $k$,
\begin{equation}
\label{smkexp}
n(k,s)\sim\frac{N_0s^{1/2}}{s^{3/2}+\lambda k^2},
\end{equation}
where $\lambda=\kappa D_b^{1/2}/2$. In this limit the surface propagator
$n(z,t)$ interestingly fulfills the time fractional diffusion equation
\cite{mekla_epl,gorenflo}
\begin{equation}
\label{tfde}
\frac{\partial^{3/2}}{\partial t^{3/2}}n(z,t)=\lambda\frac{\partial^2}{
\partial z^2}n(z,t),
\end{equation}
with the initial conditions $n(z,t=0)=N_0\delta(z)$ and $\partial n(z,t)/
\partial t\Big|_{t=0}=0$, the second defining the initial velocity field.
Here, the fractional Caputo derivative is defined via its Laplace transform
through \cite{gorenflo1,podlubny}
\begin{eqnarray}
\nonumber
\mathscr{L}\left\{\frac{\partial^{3/2}}{\partial t^{3/2}}n(z,t)\right\}&=&
s^{3/2}n(z,s)-s^{1/2}n(z,t=0)\\
&&-s^{-1/2}\left(\frac{\partial}{\partial t}n(z,t)\right)_{t=0}.
\end{eqnarray}
An equation of the form (\ref{tfde}) can be interpreted as a retarded wave
(ballistic) motion \cite{mekla_epl,meno}. We choose that the initial velocity
field $\dot{n}(z,t)\big|_{t=0}$ vanishes. It is easy to show that
Eq.~(\ref{tfde}) leads to the scaling $\langle z^2(t)\rangle\simeq t^{3/2}$
of the surface mean squared displacement.

The inverse Fourier transform of Eq.~(\ref{smkexp}) leads to
\begin{equation}
n(z,s)\sim\frac{N_0}{2\lambda^{1/2}s^{1/4}}\exp\left(-\frac{s^{3/4}}{\lambda^{
1/2}}|z|\right).
\end{equation}
Inverse Laplace transform then yields
\begin{equation}
n(z,t)\sim\frac{N_0}{2\lambda^{1/2}t^{3/4}}M\left(\zeta,\frac{3}{4}\right),
\end{equation}
where we use the abbreviation
\begin{equation}
\zeta=\frac{|z|}{\lambda^{1/2}t^{3/4}}=\sqrt{2}\frac{|z|}{\ell_C(t)}\left(
\frac{t_{\kappa}}{t}\right)^{1/4},
\end{equation}
and where $M(\zeta,\beta)$ is the Mainardi function, defined in terms of its
Laplace transform as \cite{mainardi,podlubny}
\begin{equation}
M(\zeta,\beta)=\frac{1}{2\pi i}\int_{\mathrm{Br}}\frac{d\sigma}{\sigma^{1-
\beta}}e^{\sigma-\zeta\sigma^{\beta}},\,\,\,0<\beta<1.
\end{equation}
In the tails of the distribution, i.e., in the limit $\zeta\gg1$, we may thus
employ the asymptotic form of the Mainardi function,
\begin{equation}
M\left(\frac{r}{\beta},\beta\right)\sim a(\beta)r^{(\beta-1/2)/(1-\beta)}\exp
\left(-b(\beta)r^{1/(1-\beta)}\right),
\end{equation}
for $r\to\infty$, where
\begin{equation}
a(\beta)=\frac{1}{\sqrt{2\pi(1-\beta)}},\,\,\,b(\beta)=\frac{1-\beta}{\beta}>0.
\end{equation}
We then arrive at the asymptotic form
\begin{equation}
n(z,t)\sim C_1\frac{N_0|z|}{\lambda t^{3/2}}\exp\left(-C_2\frac{z^4}{\lambda^2
t^3}\right),
\end{equation}
where $C_1$ and $C_2$ are positive constants. Thus, the
Cauchy distribution in the central part is truncated by compressed Gaussian
tails decaying as $\exp\left(-z^4/t^3\right)$ \cite{REMMM}.

\subsection{Intermediate times $t_{\kappa}\ll t\ll t_a\ll t_b$}

The range of intermediate times $t_{\kappa}\ll t\ll t_a\ll t_b$ in the Laplace
domain corresponds to $st_{\kappa}\ll 1$ while $st_a,st_b\gg1$.

\subsubsection{Surface propagator in Fourier-Laplace space}

As the characteristic time $t_{\kappa}$ does not appear in the expressions
$\Delta$ and $\Delta_1$, the limiting form (\ref{limprop}) is still valid
in this regime.

\subsubsection{Number of particles on the surface}

While Eq.~(\ref{nsurf_1}) still holds, the leading behavior of $N_s(t)$
changes, as now $st_{\kappa}\ll1$:
\begin{equation}
\label{num_int}
N_s(s)\sim\frac{N_0t_{\kappa}^{1/2}}{s^{1/2}},
\end{equation}
and thus
\begin{equation}
N_s(t)\sim\frac{N_0}{\pi^{1/2}}\frac{t_{\kappa}^{1/2}}{t^{1/2}}.
\end{equation}
In this intermediate regime the number of surface particles decays in a
square root fashion with time $t$.

\subsubsection{Surface mean squared displacement}

In a similar fashion Eq.~(\ref{surf_scl_1}) remains valid, however, as we now
encounter the limit $st_{\kappa}\ll1$ we obtain the following time dependence,
\begin{equation}
\langle z^2(s)\rangle\sim\frac{2D_st_{\kappa}}{s}+\frac{D_bt_{\kappa}^{1/2}}{
s^{3/2}}.
\end{equation}
After Laplace inversion the slow square root behavior
\begin{equation}
\label{msd_2}
\langle z^2(t)\rangle\sim2D_st_{\kappa}+\frac{2}{\pi^{1/2}}D_b\sqrt{t_{\kappa}
t}
\end{equation}
in time yields. As the number of surface particles is no longer constant,
we obtain the normalized form of the surface mean squared displacement,
\begin{equation}
\langle z^2(t)\rangle_{\mathrm{norm}}\sim2D_s\sqrt{\pi t_{\kappa}t}+2D_bt:
\end{equation}
corrected for the square root loss of surface particles to the bulk, the
normalized surface mean squared displacement exhibits normal diffusion.

\subsubsection{Surface propagator in real space}

In this intermediate time regime $st_{\kappa}\ll1$ and $st_a,st_b\gg1$ from
expression (\ref{limprop}) we obtain
\begin{equation}
n(k,s)\sim N_0\frac{t_{\kappa}^{1/2}}{\sqrt{s+D_bk^2}}.
\end{equation}
Recalling the translation theorem of the Laplace transform
\begin{equation}
f(s-a)\div e^{at}f(t),
\end{equation}
and identifying $f(s)=s^{-1/2}$, we readily find
\begin{equation}
n(k,t)\sim N_0\sqrt{\frac{t_{\kappa}}{\pi t}}\exp\Big(-D_bk^2t\Big)
\end{equation}
and thus obtain the quasi-Gaussian form
\begin{equation}
n(z,t)\sim N_0\sqrt{\frac{t_{\kappa}}{4\pi^2D_bt^2}}\exp\left(-\frac{z^2}{4D_bt}
\right).
\end{equation}
This function is not normalized, corresponding to the time evolution of the
surface particle number $N_s(t)\sim N_0\sqrt{t_{\kappa}/(\pi t)}$.

\subsection{Longer times $t_{\kappa}\ll t_a\ll t\ll t_b$}
\label{scl_longer}

In the regime of longer times $t_{\kappa}\ll t_a\ll t\ll t_b$ the corresponding
inequality in the Laplace domain reads $st_{\kappa},st_a\ll1$ while $st_b\gg1$.

\subsubsection{Surface propagator in Fourier-Laplace space}

In this limit we may take $qb\gg1$ and therefore have $I_1(qb)\to0$. The ratio
$\Delta_1/\Delta$ is therefore approximated by
\begin{equation}
\frac{\Delta_1}{\Delta}\sim\frac{K_1(qa)}{K_0(qa)},
\end{equation}
and we find the following limiting form for the surface propagator,
\begin{equation}
\label{surfprop_3}
n(k,s)\sim\frac{N_0}{\displaystyle s+k^2D_s+\kappa q\frac{K_1(qa)}{K_0(qa)}}.
\end{equation}

\subsubsection{Number of particles on the surface}

Using again the relation $N_s(s)=n(k=0,s)$ and with $q=\sqrt{s/D_b}$ at
$k=0$, we obtain
\begin{equation}
N_s(s)=\frac{N_0}{\displaystyle s+\frac{s}{\sqrt{st_{\kappa}}}\frac{K_1\left(
\sqrt{st_a}\right)}{K_0\left(\sqrt{st_a}\right)}}.
\end{equation}
We proceed to approximate the Bessel functions in this expression. For small
argument $x$,
\begin{eqnarray}
\nonumber
&&K_0(x)\approx-\left(\ln\frac{x}{2}+\gamma\right)\\
&&K_1(x)\approx\frac{1}{x},
\label{bess_exp_3}
\end{eqnarray}
where $\gamma\approx0.5772$ is Euler's constant. With $st_a,st_b\ll1$ we thus
arrive at the form
\begin{equation}
\label{n_as}
N_s(s)\sim\frac{N_0}{2}\sqrt{t_at_{\kappa}}\ln\left(\frac{4}{C^2st_a}\right),
\end{equation}
where $\ln C\equiv\gamma$. After Laplace inversion (see App.~\ref{lap_trans})
we obtain the final $1/t$ result for the number of surface particles,
\begin{equation}
\label{nsurf_3}
N_s(t)\sim\frac{N_0}{2}\sqrt{t_at_{\kappa}}\frac{1}{t}.
\end{equation}

\subsubsection{Surface mean squared displacement}

The surface mean squared displacement can be obtained from expansion of
the surface propagator (\ref{surfprop_3}) at small $k$. Some care has to
be taken to consistently expand the Bessel functions. We proceed as
follows. Since $qa\ll1$ we make use of the expansions (\ref{bess_exp_3})
and find
\begin{eqnarray}
\nonumber
n(k,s)&\sim&\frac{N_0}{\displaystyle s+k^2D_s+\frac{2\kappa}{a}\frac{1}{\ln
\left(\frac{4}{C^2(k^2a^2+st_a)}\right)}}\\
\nonumber
&\sim&\frac{N_0}{\displaystyle s+k^2D_s+\frac{2\kappa}{a}\frac{1}{\ln
\left(\frac{4}{C^2st_a}\right)-\frac{k^2a^2}{st_a}}}
\end{eqnarray}
We then expand in the denominator according to
\begin{equation}
n(k,s)\sim\frac{N_0}{\displaystyle s+k^2D_s+\frac{2\kappa}{a}\frac{1}{\left(
\frac{4}{C^2st_a}\right)}\left[1+\frac{k^2a^2}{st_a\ln\left(\frac{4}{C^2st_a}
\right)}\right]}
\end{equation}
Expansion in orders of $k$ finally leads us to
\begin{eqnarray}
\nonumber
n(k,s)&\sim&N_0\left\{
\frac{\sqrt{t_at_{\kappa}}}{2}\ln\left(\frac{4}{C^2st_a}\right)
-k^2D_b\frac{\sqrt{t_at_{\kappa}}}{2s}.\right.\\
&&\left.-k^2D_s\frac{t_at_{\kappa}}{4}\ln^2\left(\frac{4}{C^2st_a}\right)
+\mathcal{O}\left(k^4\right)\right\}.
\end{eqnarray}
From this expression we can now obtain the surface mean squared displacement
in the form
\begin{eqnarray}
\nonumber
\langle z^2(s)\rangle&=&-\left.\frac{\partial^2n(k,s)}{\partial k^2}\right|_{
k=0}\\
&\sim&D_b\frac{\sqrt{t_at_{\kappa}}}{s}+D_s\frac{t_at_{\kappa}}{2}\ln^2\left(
\frac{C^2t_a}{4}s\right).
\end{eqnarray}
Using the asymptotic Laplace transform pair (compare App.~\ref{lap_trans})
\begin{equation}
\label{lap_pair}
\ln^2(As)\div\frac{2}{t}\ln\left(\frac{Ct}{A}\right)
\end{equation}
we obtain the surface mean squared displacement
\begin{equation}
\label{msd_3}
\langle z^2(t)\rangle\sim D_s\frac{t_at_{\kappa}}{t}\ln\left(\frac{4t}{Ct_a}
\right)+D_b\sqrt{t_at_{\kappa}}.
\end{equation}
Normalized by the associated time evolution of the number of surface
particles the normalized surface mean squared displacement becomes
\begin{equation}
\langle z^2(t)\rangle_{\mathrm{norm}}\sim2D_s\sqrt{t_at_{\kappa}}\ln\left(
\frac{4t}{Ct_a}\right)+2D_bt.
\end{equation}
Again, this result is quite remarkable: the surface mean squared displacement
reaches a plateau value in this regime. In absence of the outer cylinder this
is the terminal behavior, reflecting the balance of ever increasing surface
displacement due to long bulk excursions, and the continuing escape of surface
particles to the bulk. Normalized to the time evolution of these surface
particles we find a linear growth of the surface mean squared displacement.

\subsubsection{Surface propagator in real space}

In contrast to the previous two regimes, here the value of $qa$ acquires
values smaller and larger than 1. In the tails of the propagator ($z\gg 1$)
we expand
\begin{equation}
K_0(qa)\sim-\ln(qa)\,\,\,\mbox{and}\,\,\,K_1(qa)\sim1/(qa).
\end{equation}
Therefore we can express the propagator as
\begin{equation}
n(k,s)\sim\frac{N_0}{\displaystyle s+\frac{\kappa}{a\ln(1/[qa])}}
\end{equation}
We further approximate this expression to obtain the logarithmic form
\begin{equation}
n(k,s)\sim N_0\frac{a}{\kappa}\ln\frac{1}{qa}=-N_0\frac{a}{2\kappa}\ln\left(
a^2k^2+st_a\right).
\end{equation}
That this seemingly harsh approximation makes sense can be seen by evaluating
the surface particle number $N_s(t)=N_0n(k=0,s)\sim\frac{1}{2}N_0\sqrt{t_at_{
\kappa}}\ln(1/[st_a])$ leading to $N_s(t)\sim N_0\sqrt{t_at_{\kappa}}/(2t)$,
matching our previous result (\ref{nsurf_3}).

Formally we can now write
\begin{equation}
\label{formal1}
n(k,t)=-\frac{N_0a}{2\kappa}\int_{\mathrm{Br}}e^{st}\ln\Big(st_a+k^2a^2\Big)
\frac{ds}{2\pi i}
\end{equation}
where the integral index Br indicates the Bromwich curve for the Laplace
inversion. Following App.~\ref{lap_trans}, we obtain
\begin{equation}
n(k,t)=\frac{N_0a}{2\kappa t}\exp\left(-k^2D_bt\right).
\label{formal2}
\end{equation}
Inverse Fourier transformation delivers the Gaussian result
\begin{equation}
n(z,t)\sim\frac{N_0a}{2\kappa t}\frac{1}{\sqrt{4\pi D_bt}}\exp\left(
-\frac{z^2}{4D_bt}\right)
\end{equation}
with varying normalization.

\subsection{Long times $t_{\kappa}\ll t_a\ll t_b\ll t$}
\label{scl_long}

In this final regime the outer cylinder becomes dominant, and the inequalities
$t_{\kappa}\ll t_a\ll t_b\ll t$ correspond to $st_{\kappa}, st_a,st_b\ll 1$ in
terms of the associated Laplace variable.

\subsubsection{Surface propagator in Fourier-Laplace space}

In this long time regime we start with the original expression (\ref{nsol_fl})
of the surface propagator in Fourier-Laplace space, and take the appropriate
limits for the number of surface particles and the surface mean squared
displacement separately.

\subsubsection{Number of particles on the surface}

From Eq.~(\ref{nsol_fl}) we directly obtain in the $k=0$ limit
\begin{equation}
\label{napprox_4}
N_s(s)=\frac{N_0}{\displaystyle s+\kappa\sqrt{\frac{s}{D_b}}\frac{\Delta_1
(0,s)}{\Delta(0,s)}}.
\end{equation}
To calculate the approximations for $\Delta_1(0,s)$ and $\Delta(0,s)$ we
employ the following small argument expansions of the Bessel functions:
\begin{eqnarray}
\nonumber
&&I_0(x)\approx1\\
\nonumber
&&K_0(x)\approx-\gamma-\ln\frac{x}{2}\\
\nonumber
&&I_1(x)\approx\frac{x}{2}\\
&&K_1(x)\approx\frac{1}{x}.
\label{bessapprox_4}
\end{eqnarray}
Then we find
\begin{eqnarray}
\nonumber
\Delta_1(0,s)&\approx&\frac{1}{2}\sqrt{\frac{t_b}{t_a}}\\
\Delta(0,s)&\approx&\frac{1}{\sqrt{st_b}}.
\end{eqnarray}
Plugging these expansions into expression (\ref{napprox_4}) we get
\begin{equation}
\label{ex_p}
N_s(s)\sim\frac{N_0}{\displaystyle s+s\frac{t_b}{2\sqrt{t_at_{\kappa}}}}\sim
\frac{2N_0\sqrt{t_at_{\kappa}}}{t_bs},
\end{equation}
and therefore
\begin{equation}
\label{nl_4}
N_s(t)\sim\frac{2N_0\sqrt{t_at_{\kappa}}}{t_b}.
\end{equation}
At long times the system reaches a stationary state due to the confinement
by the outer cylinder.

\subsubsection{Surface mean squared displacement}

Again we start with the full surface propagator in Fourier-Laplace space,
Eq.~(\ref{nsol_fl}), and this time expand it around $k=0$. Since $st_a,st_b
\ll1$ we have $qa,qb\ll1$. To calculate $\Delta_1(k,s)$ and $\Delta(k,s)$
at $k\to0$ we use the approximations (\ref{bessapprox_4}). Then,
\begin{eqnarray}
\nonumber
&&\Delta_1(k,s)\approx\frac{1}{2}\sqrt{\frac{t_b}{t_a}},\\
&&\Delta(k,s)\approx\frac{1}{qb}.
\end{eqnarray}
Inserting into the propagator (\ref{nsol_fl}) delivers the approximation
\begin{eqnarray}
\nonumber
n(k,s)&\sim&\frac{1}{\displaystyle s+k^2D_s+\frac{\kappa b^2}{2a}q^2}\\
&\sim&\frac{1}{\displaystyle\frac{st_b}{2\sqrt{t_at_{\kappa}}}+k^2\left(
D_s+D_b\frac{t_b}{2\sqrt{t_at_{\kappa}}}\right)}.
\end{eqnarray}
We expand this expression in powers of $k$, obtaining
\begin{equation}
n(k,s)\sim\frac{2\sqrt{t_at_{\kappa}}}{st_b}-k^2\frac{4t_at_{\kappa}}{s^2t_b^2}
\left(D_s+\frac{t_b}{2\sqrt{t_at_{\kappa}}}D_b\right)+\mathcal{O}\left(k^4
\right).
\end{equation}
For the surface mean squared displacement we therefore have that
\begin{eqnarray}
\nonumber
\langle z^2(s)\rangle&=&-\left.\frac{\partial^2n(k,s)}{\partial k^2}\right|_{
k=0}\\
&\sim&\frac{8t_at_{\kappa}}{s^2t_b^2}D_s+4D_b\frac{\sqrt{t_at_{\kappa}}}{s^2
t_b},
\end{eqnarray}
and finally
\begin{equation}
\label{msd_4}
\langle z^2(t)\rangle\sim\frac{8t_at_{\kappa}}{t_b^2}D_st+4D_b\frac{\sqrt{t_a
t_{\kappa}}}{t_b}t.
\end{equation}
At long times the stationary process causes a normal effective surface
diffusion. The normalized surface mean squared displacement attains the
form
\begin{equation}
\langle z^2(t)\rangle_{\mathrm{norm}}\sim\frac{4\sqrt{t_at_{\kappa}}}{t_b}
D_st+2D_bt.
\end{equation}
Without surface diffusion we therefore observe normal linear diffusion with
with the bulk diffusivity $D_b$. In the presence of surface diffusion we
have a correction proportional to $D_s$. Given that $t_b\gg\sqrt{t_at_{
\kappa}}$, the amplitude of this surface contribution is small.

\subsubsection{Surface propagator in real space}

In this long time regime we consider the tails of the propagator such that
$z\gg b$ and $qa,qb\ll 1$. The Bessel functions are thus approximated by
Eqs.~(\ref{bessapprox_4}); therefore,
\begin{equation}
\label{long_lim_prop}
n(k,s)\sim\frac{N_0}{s+\kappa b^2q^2/(2a)}=\frac{N_0}{s+\kappa b^2\left(k^2
+s/D_b\right)/(2a)}.
\end{equation}
Since
\begin{equation}
\frac{\kappa b^2}{2aD_b}=\frac{t_b}{2\sqrt{t_at_{\kappa}}}\gg 1
\end{equation}
we may simplify Eq.~(\ref{long_lim_prop}) to
\begin{equation}
n(k,s)\sim\frac{2N_0\sqrt{t_at_{\kappa}}}{t_b}\frac{1}{s+D_bk^2}.
\end{equation}
The propagator consequently assumes the Gaussian shape
\begin{equation}
n(z,t)\sim\frac{2N_0\sqrt{t_at_{\kappa}}}{t_b}\frac{1}{\sqrt{4\pi D_bt}}\exp
\left(-\frac{z^2}{4D_bt}\right).
\end{equation}
The prefactor $2\sqrt{t_at_{\kappa}}/t_b$ reflects the probability that a
certain portion of the particles is desorbed from the cylinder surface.

\section{Explicit calculations: intermediate coupling limit}
\label{sec_icl}

We now turn to the case of intermediate coupling defined by $t_a\ll t_{
\kappa}\ll t_b$.

\subsection{Short times $t\ll t_a\ll t_{\kappa}\gg t_b$}

In this limit corresponding to $st_a,st_{\kappa},st_b\ll1$ we obtain the
same results as in the matching limit of the strong coupling regime, compare
Sec.~\ref{scl_short}.

\subsection{Intermediate times $t_a\ll t\ll t_{\kappa}\ll t_b$}

This limit in Laplace space corresponds to the inequalities $st_a\ll 1$
and $st_{\kappa},st_b\gg 1$.

\subsubsection{Number of particles on the surface}

In Laplace space we start from the exact expression for $N_s(s)=n(k=0,s)$,
\begin{equation}
\label{nlap_int1}
N_s(s)=\frac{N_0}{\displaystyle s+\kappa\sqrt{\frac{s}{D_b}}\frac{\Delta_1(0,
s)}{\Delta(0,s)}},
\end{equation}
where $\Delta_1(k,s)$ and $\Delta(k,s)$ are defined in Eqs.~(\ref{delta1_abb})
and (\ref{delta_abb}). With the asymptotic expansions of the modified Bessel
functions for small argument summarized in Eq.~(\ref{bessapprox_4}), as well
as with the large argument asymptotics (\ref{bessel_moreexp}) we obtain from
Eq.~(\ref{nlap_int1}) the result
\begin{equation}
N_s(s)=\frac{N_0t_c}{\displaystyle st_c+\frac{2}{\ln\left(4/[C^2st_a]\right)}},
\end{equation}
where we introduced the new time scale $t_c\equiv\sqrt{t_{\kappa}t_a}$ which
fulfills the inequality $t_a<t_c<t_{\kappa}$.

Let us first regard the subregime $t_a\ll t\ll t_c\ll t_{\kappa}$. If these
inequalities are fulfilled, we may neglect the logarithmic term in the
denominator, and find
\begin{equation}
N_s(s)\sim\frac{N_0}{s},
\end{equation}
i.e., the number of particles on the surface still remains constant to
leading order:
\begin{equation}
N_s(t)\sim N_0.
\end{equation}

The range $t_a\ll t_c\ll t\ll t_{\kappa}$ is difficult to estimate, as now
the term linear in $s$ and the logarithmic term in the denominator are of
comparable order. As can be seen in Fig.~\ref{num}, in the interval from
$t_c$ to $t_{\kappa}$ the number $N_s(t)$ of surface particles describes a
quite complicated turnover from the
persisting initial condition $N_0$ to the $1/t$ behavior in the following
regime $t_{\kappa}\ll t\ll t_b$. The prominent shoulder visible in the
double-logarithmic plot propagates to the behavior of the normalized
mean squared displacement discussed below.

\subsubsection{Surface mean squared displacement}

We start from expression (\ref{nsol_fl}) for the Fourier-Laplace transform
of the surface propagator. To determine the associated mean squared
displacement we will need the small wave number approximation. Since
$st_a\ll1$ and $st_b\gg1$ we have $qa\ll1$ and $qb\gg1$. With the
definitions (\ref{delta1_abb}) and (\ref{delta_abb}) and with the
asymptotic expansions of the modified Bessel functions we find after a few
steps
\begin{equation}
\Delta(k,s)\sim\frac{\exp(qb)}{(2\pi)^{1/2}\left(qb\right)^{1/2}}\ln\left(
\frac{2}{Cqa}\right)
\end{equation}
and
\begin{equation}
\Delta_1(k,s)\sim\frac{1}{qa}\frac{\exp(qb)}{(2\pi)^{1/2}\left(qb\right)^{
1/2}}.
\end{equation}
Thus, the following approximation
\begin{equation}
\frac{\Delta_1(k,s)}{\Delta(k,s)}\sim\frac{1}{\displaystyle qa\ln\left(
\frac{2}{Cqa}\right)}
\end{equation}
yields for the ratio $\Delta_1/\Delta$, and the surface propagator becomes
\begin{equation}
\label{int2_propfl}
n(k,s)\sim\frac{N_0}{\displaystyle s+k^2D_s+\frac{\kappa}{a}\frac{1}{\ln\left(2/
[Cqa]\right)}}.
\end{equation}
At $k\to0$ we expand the logarithm as follows:
\begin{eqnarray}
\nonumber
\ln\left(\frac{2}{Cqa}\right)&=&\ln\left(\frac{2}{C\sqrt{k^2a^2+st_a}}\right)\\
\nonumber
&=&\frac{1}{2}\ln\left(\frac{4}{C^2\left(st_a+k^2a^2\right)}\right)\\
&\sim&\frac{1}{2}\ln\left(\frac{4}{C^2st_a}\right)-\frac{k^2a^2}{2st_a}.
\end{eqnarray}
Therefore
\begin{equation}
\frac{1}{\displaystyle \ln\left(\frac{2}{Cqa}\right)}\sim\frac{2}{
\displaystyle \ln\left(\frac{4}{C^2st_a}\right)}\left[1+\frac{k^2a^2}{st_a}
\frac{1}{\displaystyle\ln\left(\frac{4}{C^2st_a}\right)}\right].
\end{equation}
Plugging this expansion into expression (\ref{int2_propfl}) we obtain
\begin{widetext}
\begin{equation}
n(k,s)\sim\frac{N_0}{\displaystyle s+\frac{2\kappa}{a}\frac{1}{\ln\left(4/[C^2
st_a]\right)}+k^2\left[D_s+\frac{2\kappa}{a}\frac{a^2}{st_a}\frac{1}{\ln^2
\left(4/[C^2st_a]\right)}\right]}.
\end{equation}
This can be rephrased in the form
\begin{equation}
n(k,s)=N_0\left[
\frac{1}{\displaystyle s+\frac{2\kappa}{a}\frac{1}{\ln\left(4/[C^2st_a]
\right)}}-k^2\frac{\displaystyle D_s+\frac{2D_b}{s\sqrt{t_{\kappa}t_a}}\frac{
1}{\ln^2\left(4/[C^2st_a]\right)}}{\displaystyle\left[s+\frac{2\kappa}{a}
\frac{1}{\ln\left(4/[C^2st_a]\right)}\right]^2}\right].
\end{equation}
\end{widetext}
Here we again consider the subregime $t_a\ll t\ll t_c\ll t_{\kappa}$ for which
in the limit $k\to0$
\begin{equation}
n(k,s)\sim N_o\left[
\frac{1}{s}-\frac{k^2D_s}{s^2}-k^2\frac{2D_b}{s^3t_c}\frac{1}{
\displaystyle\ln^2\left(\frac{4}{C^2st_a}\right)}\right],
\end{equation}
and thus
\begin{equation}
\label{aao}
\langle z^2(s)\rangle\sim\frac{2D_s}{s^2}+\frac{4D_b}{t_c}\frac{1}{
\displaystyle s^3\ln^2\left(\frac{4}{C^2st_a}\right)}.
\end{equation}
After Laplace inversion we ultimately find
\begin{equation}
\label{aao1}
\langle z^2(t)\rangle\sim2D_st+\frac{2D_bt^2}{\displaystyle t_c\ln^2\left(
\frac{4t}{C^2t_a}\right)}.
\end{equation}
According to above findings this is also the result for the normalized
surface mean squared displacement.

Analogous to what was said above, the following subregime $t_c\ll t\ll t_{
\kappa}$ is difficult to estimate analytically, and we refer to the numerical
result shown in Figs.~\ref{msd}. While for the surface mean squared
displacement one can see a slight increase in the slope compared to the
linear behavior shown by the guiding line, for the normalized analog
we see a distinct increase in the slope after $t_c$.

\subsection{Times longer than $t_a\ll t_{\kappa}$}

At times longer than the characteristic scale $t_{\kappa}$ our results again
correspond to those of the strong coupling limit, see Secs.~\ref{scl_longer}
and \ref{scl_long}.

\section{First passage statistics}

In this Section we address the problem of the first passage time statistics
in our geometry, that is, the time it takes a particle starting at some point
in between the two cylinders to reach the inner cylinder. As before we neglect
the dependence on the polar angle $\theta$ in our description. The relevant
probability density is therefore $P(r,z,t)$. Then $2\pi rP(r,z,t)drdz$ gives us
the probability that the particle at time $t$ is in the range $(r\ldots r
+dr,z\ldots z+dz)$. The initial distribution is smeared out on a circle of
radius $r_0$ in the plane $z=0$,
\begin{equation}
P(r,z,t)\Big|_{t=0}=\frac{1}{2\pi r_0}\delta(r-r_0)\delta(z).
\end{equation}
Here the factor $1/2\pi r_0$ appears because of the normalization of the
initial density,
\begin{equation}
\int_a^brdr\int_0^{2\pi}d\theta\int_{-\infty}^{\infty}dz\,P(r,z,t)\Big|_{
t=0}=1.
\end{equation}
The time evolution of $P(r,z,t)$ is given by the diffusion equation
\begin{equation}
\label{fpt_de}
\frac{\partial}{\partial t}P(\mathbf{r},z,t)=D_b\nabla^2P(\mathbf{r},z,t),
\end{equation}
valid for radii $a\le r\le b$ and on the entire cylinder axis, $-\infty<z<
\infty$. The Laplace operator in polar-symmetric cylindrical coordinates is
\begin{equation}
\nabla^2=\frac{1}{r}\frac{\partial}{\partial r}\left(r\frac{\partial}{\partial
r}\right)+\frac{\partial^2}{\partial z^2}.
\end{equation}
In the calculation of the first passage dynamics we impose an absorbing
boundary condition at $r=a$ such that
\begin{equation}
P(r,z,t)\Big|_{r=a}=0,
\end{equation}
while at the outer cylinder we keep the reflecting boundary condition
\begin{equation}
\frac{\partial}{\partial r}P(r,z,t)\Big|_{r=b}=0.
\end{equation}
The result for the probability density is
\begin{widetext}
\begin{equation}
P(r,k,s)=\frac{1}{2\pi D_b}\frac{I_1(qb)K_0(qr_0)+K_1(qb)I_0(qr_0)}{I_0(qa)
K_1(qb)+K_0(qa)I_1(qb)}\Big(K_0(qa)I_0(qr)-I_0(qa)K_0(qr)\Big).
\label{sol}
\end{equation}
\end{widetext}
as calculated in Appendix \ref{app_fpt}.

\subsection{First passage time density for times $t\ll t_b$}

We first investigate the case when the outer cylinder is remote, that is,
$t\ll t_b$. To this end we set $b\to\infty$. In Eq.~(\ref{sol}) this means
that $I_1(qb)\to\infty$ and $K_1(qb)\to0$ such that
\begin{equation}
P(r,k,s)=\frac{K_0(qr_0)\Big(K_0(qa)I_0(qr)-I_0(qa)K_0(qr)\Big)}
{2\pi D_bK_0(qa)}.
\label{fpt_res}
\end{equation}

The probability density function for the first passage time is given
by the radial flux through
\begin{equation}
\wp(t)=2\pi a\int_{-\infty}^{\infty}D_b\left.\frac{\partial P(r,z,t)}{
\partial r}\right|_{r=a}dz,
\end{equation}
compare also Ref.~\cite{berg}. Its Laplace transform reads
\begin{equation}
\wp(s)=2\pi aD_b\left.\frac{\partial P(r,k,s)}{\partial r}\right|_{r=a,k=0}.
\label{lap_flux}
\end{equation}
where the integral over the cylinder axis $z$ has been replaced by the
zeroth Fourier mode. Inserting Eq.~(\ref{fpt_res}),
\begin{eqnarray}
\nonumber
\wp(s)&=&qa\frac{K_0(qr_0)}{K_0(qa)}\Big\{I_0(qa)K_1(qa)\\
\nonumber
&&\hspace*{2.4cm}+K_0(qa)I_1(qa)\Big\}_{q=\sqrt{s/D_b}},\\
\nonumber
&=&\sqrt{st_a}\frac{K_0\left(\sqrt{st_0}\right)}{K_0\left(\sqrt{st_a}\right)}
\Big\{I_0\left(\sqrt{st_a}\right)K_1\left(\sqrt{st_a}\right)\\
\nonumber
&&\hspace*{2.4cm}+K_0\left(\sqrt{st_a}\right)I_1\left(\sqrt{st_a}\right)
\Big\}\\
&=&\frac{K_0\left(\sqrt{st_0}\right)}{K_0\left(\sqrt{st_a}\right)}.
\label{res_lev}
\end{eqnarray}
Here we defined the diffusion time
\begin{equation}
t_0=\frac{r_0^2}{D_b}.
\end{equation}
Expression (\ref{res_lev}) recovers a result in Ref.~\cite{levitz}.

To evaluate this result we need the more subtle expansion of the Bessel
functions \cite{abramowitz}
\begin{eqnarray}
\nonumber
K_0(z)&\sim&-I_0(z)\left\{\gamma+\ln\left(\frac{z}{2}\right)\right\}+\frac{
z^2}{4}\\
&\sim&-\ln z+C+\mathcal{O}\left(z^2\right).
\end{eqnarray}
Here $\gamma\approx0.5772$ is Euler's constant such that $C=\ln2-\gamma>0$.
We therefore find for the Laplace image of the first passage time density
\begin{eqnarray}
\nonumber
\wp(s)&\sim&\frac{\ln(1/[st_0])+2C}{\ln(1/[st_a])+2C}\\
\nonumber
&\sim&\frac{\ln(1/[st_0])}{\ln(1/[st_a])}\left[1+\frac{2C}{\ln(1/[st_0])}
\right]\left[1-\frac{2C}{\ln(1/[st_a])}\right]\\
\nonumber
&\sim&\frac{\ln\left(\frac{1}{st_a}\frac{t_a}{t_0}\right)}{\ln\left(\frac{1}{
st_a}\right)}\\
&\sim&1-\frac{\ln(t_0/t_a)}{\ln(1/[st_a])}.
\end{eqnarray}
Substituting for $t_0$ we obtain
\begin{equation}
\wp(s)\sim1-2\frac{\ln(r_0/a)}{\ln(1/[st_a])}.
\end{equation}
The Laplace inversion based on Tauberian theorems for slowly varying functions
\cite{havlin} finally delivers the desired result
\begin{equation}
\wp(t)\simeq2\frac{\ln(r_0/a)}{t\ln^2(t/t_a)}.
\end{equation}
This expansion is valid in the range $t\gg t_a$.
We therefore obtain a very subtle probability density, in which the logarithm
ensures normalizability, however, not even fractional moments $\langle t^q
\rangle$ with $q>0$ exist. This extremely shallow first passage time density
is characteristic for the cylindrical problem. We note that in the limit
$r_0=a$ we recover $\wp(t)=\delta(t)$, as it should be.

\subsection{First passage time density for times $t\gg t_b$}

At times $t\gg t_b$ the outer cylinder comes into play.
To assess the behavior of the first passage in this regime
we insert the full solution (\ref{sol}) into the equation (\ref{lap_flux})
for the flux, finding
\begin{equation}
\wp(s)=\frac{I_1\left(\sqrt{st_b}\right)K_0\left(\sqrt{st_0}\right)+K_1
\left(\sqrt{st_b}\right)I_0\left(\sqrt{st_0}\right)}{I_1\left(\sqrt{st_b}
\right)K_0\left(\sqrt{st_a}\right)+K_1\left(\sqrt{st_b}\right)I_0
\left(\sqrt{st_a}\right)}.
\end{equation}
At $st_a<st_0<st_b\ll1$ we use the following expansions for the Bessel
functions
\begin{eqnarray}
\nonumber
I_1\left(\sqrt{st_b}\right)&\sim&\frac{1}{2}\sqrt{st_b},\\
\nonumber
K_1\left(\sqrt{st_b}\right)&\sim&\frac{1}{\sqrt{st_b}},\\
\nonumber
I_0\left(\sqrt{st_{a/0}}\right)&\sim&1,\\
K_0\left(\sqrt{st_{a/0}}\right)&\sim&-\ln\sqrt{st_{a/0}}.
\end{eqnarray}
This leads us to
\begin{eqnarray}
\nonumber
\wp(s)&\sim&\frac{1-st_b\ln(st_0)/4}{1-st_b\ln(st_a)/4}\\
\nonumber
&\sim&1-\frac{st_b}{4}\ln\left(\frac{t_0}{t_a}\right)\\
&\sim&1-\langle t\rangle s.
\end{eqnarray}
At long times $t\gg t_b$ we find a finite mean first passage time
\begin{equation}
\langle t\rangle=\frac{b^2}{2D_b}\ln\left(\frac{r_0}{a}\right),
\end{equation}
as it should be in this stationary regime.

\begin{figure*}
\includegraphics[width=8.8cm]{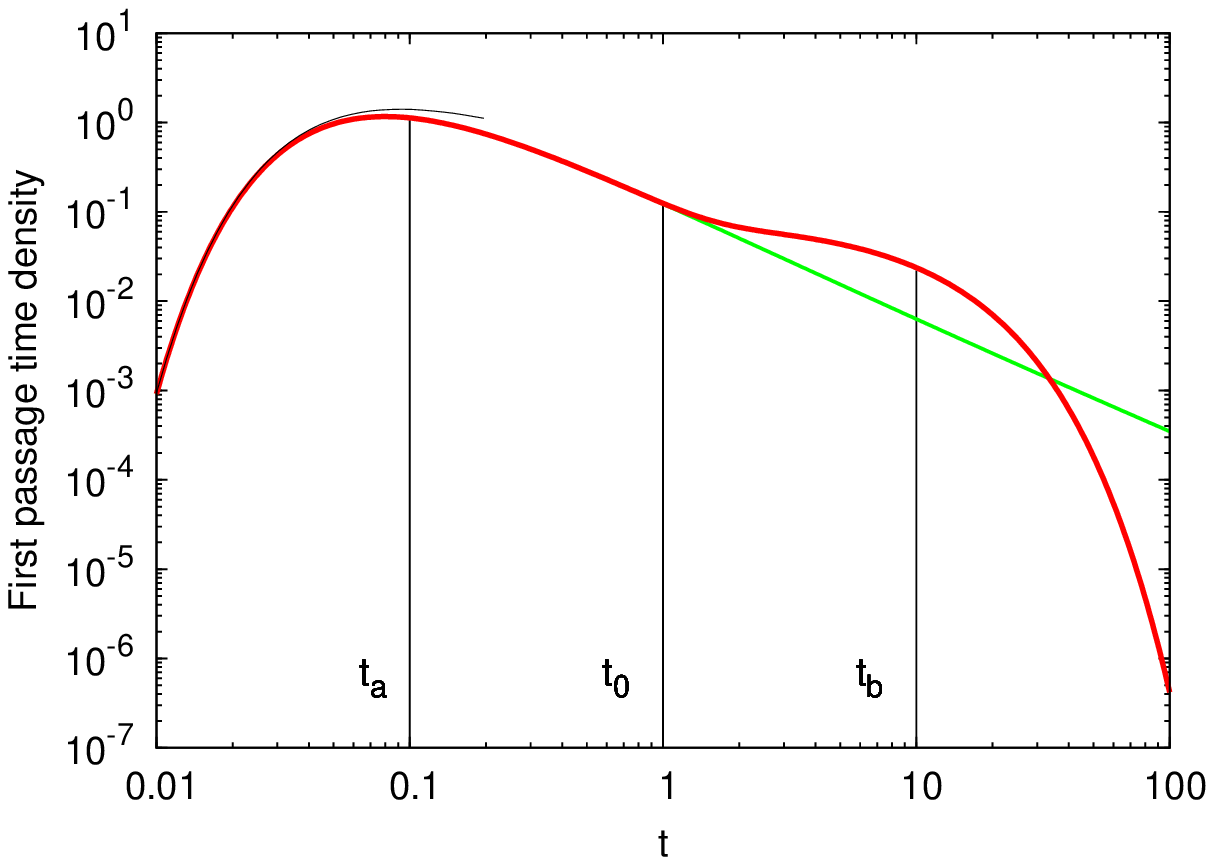}
\includegraphics[width=8.8cm]{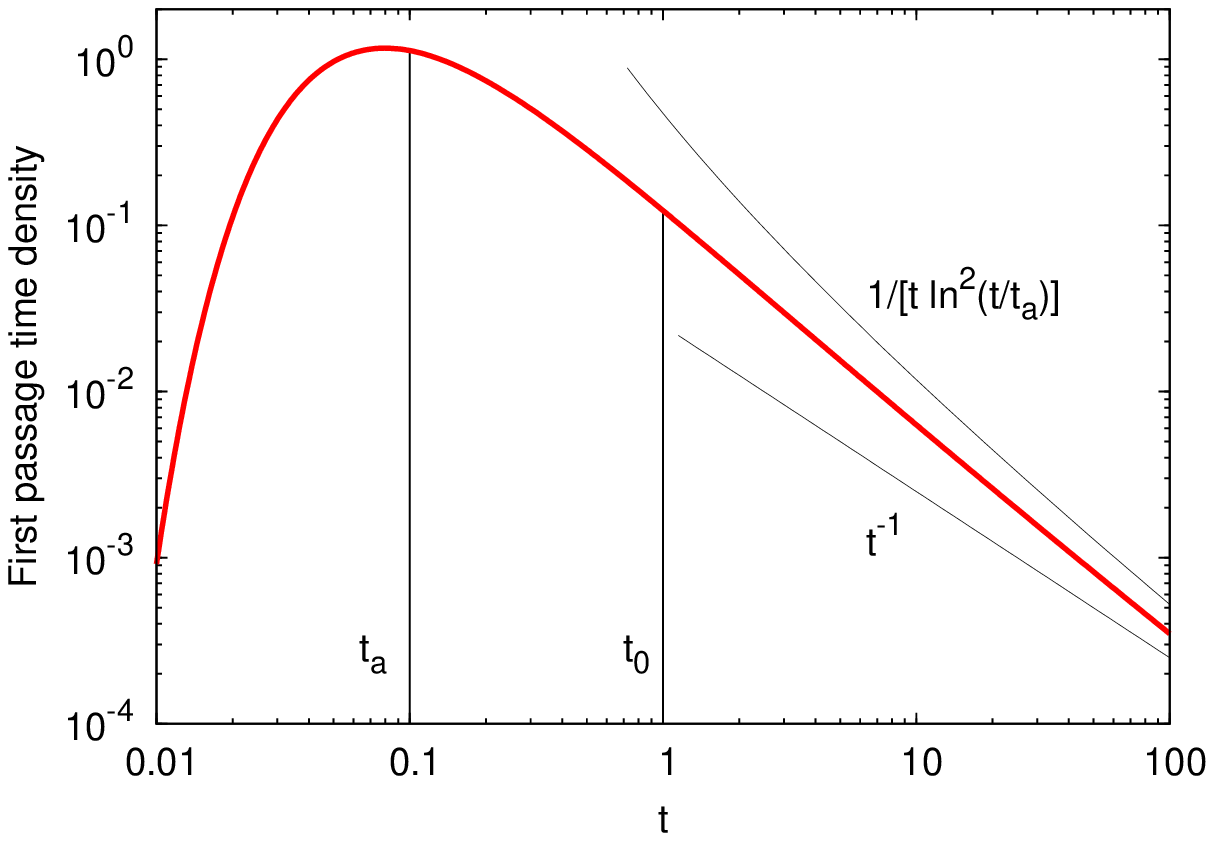}
\caption{First passage time density $\wp(t)$. Left: The outer cylinder is
present, eventually causing an exponential-like steep decay of $\wp(t)$,
compared to the unbounded behavior.
The black line shows the short time approximation (\ref{fpt_vshort}).
Right: Unbounded problem showing the long-time asymptotics. The inverse
square-logarithmic term matters, as can be seen from comparison with the shown
asymptotic behaviors.}
\end{figure*}

\subsection{First passage time density at very short times $t\ll t_a$}

We conclude our discussion of the first passage time density with the case
of very short times, $t\ll t_a$. In this regime a particle
starting close to the inner cylinder surface does not yet
feel the cylindrical geometry and we would naively expect that the first
passage is given by the one-dimensional L{\'e}vy-Smirnov form.

As we may neglect the outer cylinder, we start from result (\ref{fpt_res}).
We expand this result in inverse powers of $\sqrt{s}$ and then perform a
term-wise inverse Laplace transform. With
\begin{equation}
K_{\nu}(z)=\sqrt{\frac{\pi}{2z}}e^{-z}\left\{1+\frac{4\nu^2-1}{8z}+\mathcal{O}
\left(\frac{1}{z^2}\right)\right\},
\end{equation}
we find that
\begin{eqnarray}
\nonumber
\frac{K_0\left(\sqrt{st_0}\right)}{K_0\left(\sqrt{st_a}\right)}&\sim&
\left(\frac{t_a}{t_0}\right)^{1/4}\exp\left(-\sqrt{s}\Big[\sqrt{t_0}-
\sqrt{t_a}\Big]\right)\\
\nonumber
&&\times\left\{1+\frac{1}{8}\left(\frac{1}{\sqrt{st_a}}-\frac{1}{\sqrt{
st_0}}\right)\right\}\\
\nonumber
&\sim&\left(\frac{a}{r_0}\right)^{1/2}\exp\left(-\sqrt{s}\frac{r_0-a}{
\sqrt{D_b}}\right)\\
&&\times\left\{1+\frac{\sqrt{D_b}}{8\sqrt{s}}\frac{r_0-a}{ar_0}\right\}.
\end{eqnarray}
We are thus led to the inverse Laplace transform
\begin{eqnarray}
\nonumber
\wp(t)&\sim&\left(\frac{a}{r_0}\right)^{1/2}\frac{r_0-a}{\sqrt{D_b}}\frac{1}{
2\sqrt{\pi t^3}}\exp\left(-\frac{(r_0-a)^2}{4D_bt}\right)\\
&&\hspace*{-0.8cm}
+\left(\frac{a}{r_0}\right)^{1/2}\frac{r_0-a}{ar_0}\frac{\sqrt{D_b}}{8\sqrt{
\pi t}}\exp\left(-\frac{(r_0-a)^2}{4D_bt}\right).
\end{eqnarray}
Reorganizing this expression we find
\begin{eqnarray}
\nonumber
\wp(t)&\sim&\left(\frac{a}{r_0}\right)^{1/2}\frac{r_0-a}{\sqrt{4\pi D_bt^3}}
\exp\left(-\frac{(r_0-a)^2}{4D_bt}\right)\\
&&\times\left\{1+\frac{D_bt}{4ar_0}+\ldots\right\}.
\label{fpt_vshort}
\end{eqnarray}
At very short times the first passage time density indeed coincides with the
one-dimensional limit, reweighted by the ratio $\sqrt{a/r_0}$. Note that if
we keep the distance $\Delta=r_0-a$ fixed but let both $r_0$ and $a$ tend to
infinity, we recover the result for a flat surface,
\begin{equation}
\wp_{\mathrm{flat}}(t)=\frac{\Delta}{\sqrt{4\pi D_bt^3}}\exp\left(-\frac{
\Delta^2}{4D_bt}\right),
\end{equation}
i.e., the well-known L{\'e}vy-Smirnov distribution.

\section{Discussion}

We established an exact approach to BMSD along a reactive
cylindrical surface revealing four distinct diffusion regimes. In
particular our formalism provides a stringent derivation of the transient
superdiffusion discussed earlier \emph{and\/} explicitly quantifies the
transition to other regimes. Notably we revealed a saturation regime for the
MSD along the cylinder that becomes relevant at times above which the diffusing
particle feels the curvature of the cylinder surface ($t_a$). This behavior,
caused by the cylindrical geometry,
stems from an interesting balance between a net flux of particles into the
bulk and the fact that particles with a longer return time also lead to an
increased effective surface relocation. In absence of an outer cylinder the
saturation is terminal, while in its presence the MSD along the cylinder
returns to a
linear growth in time. This observation will be important in future models
of BMSD around cylinders and particularly for the interpretation of
experimental data obtained for BMSD systems. We note that in the proper
limit $a\to\infty$ the previous results for a planar surface are recovered.
Relaxing the strong coupling condition we demonstrated the existence of an
almost ballistic BMSD behavior, a case that might be relevant
for transport along thin cylinders such as DNA.

In Ref.~\cite{levitz} it was shown that the scaling behavior in the
regimes below and above $t_a$ can be probed experimentally by NMR methods
measuring the BMSD of water molecules along imogolite nanorods over three
orders of magnitude in frequency space. For larger molecules such as a
protein of approximate diameter 5 nm we observe a diffusivity of $10^{-6}
\mathrm{cm}^2/\mathrm{sec}$ such that for instance the saturation plateau
around a bacillus cell (radius 1/2 $\mu$m) sets in at around $t_a=2.5$ msec
which might give rise to interesting consequences for the material exchange
around such cells. In general, the relevance of the individual regimes will
crucially depend on the scales of the surface radius and the diffusing
particle (and therefore its diffusivity). It was discussed previously that
even the superdiffusive short-term behavior may become relevant
\cite{bychuk,bychuk1,fatkullin}. In general, in a given system the separation
between the various scaling regimes may not be sharp. Moreover typically a
single experimental technique will not be able to probe all regimes. It is
therefore vital to have available a solution for the entire BMSD problem.

\begin{acknowledgments}
Support from the Deutsche Forschungsgemeinschaft, the European Commission
(grant MC-IIF 219966), and the Academy of Finland (FiDiPro scheme) are
gratefully acknowledged.
\end{acknowledgments}

\begin{appendix}

\section{Derivation of the reactive boundary condition for a planar surface}
\label{reactive_bc}

\begin{figure}
\includegraphics[width=8.8cm]{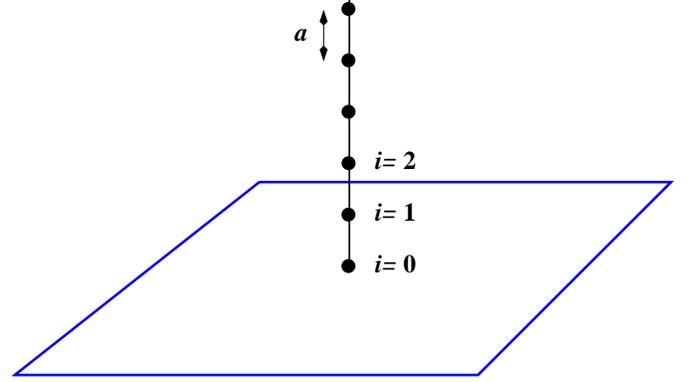}
\caption{Schematic of the random walk picture of the surface-bulk exchange.
In our derivation we pass from the discrete lattice with spacing $a$ to the
continuous variable $\rho$. The surface corresponds to $\rho=0$.}
\label{schematic}
\end{figure}

We start with a derivation of the coupling between surface and bulk in a
discrete random walk process along the $\rho$ coordinate perpendicular to
the surface, as specified in Fig.~\ref{schematic}. Let $N_i$ with
$i=1,2,\ldots$ denote the number of particles at site $i$ of this
one-dimensional lattice with spacing $a$. The number of particles on
the surface at lattice site $i=0$ are termed $\mathcal{N}_0$. The
exchange of particles is possible only via nearest neighbor jumps,
each characterized by the waiting time $\tau$. For the exchange between
the surface and site $i=1$ we then have the following law
\begin{equation}
\label{exch}
\frac{d\mathcal{N}_0(t)}{dt}=\frac{1}{2\tau}N_1-\frac{1}{\tau_{\mathrm{des}}}
\mathcal{N}_0,
\end{equation}
where $\tau_{\mathrm{des}}$ is the characteristic time for desorption
from the surface. The bulk sites are governed by equations of the form
\begin{eqnarray}
\nonumber
\frac{dN_1(t)}{dt}&=&\frac{1}{\tau_{\mathrm{des}}}\mathcal{N}_0-\frac{1}{\tau}
N_1+\frac{1}{2\tau}N_2,\\
\frac{dN_2(t)}{dt}&=&\frac{1}{2\tau}N_1+\frac{1}{2\tau}N_3-\frac{1}{\tau}N_2,
\label{exch1}
\end{eqnarray}
etc. Let us define the number of ``bulk'' particles at the surface site $i=0$
through
\begin{equation}
\label{exch4}
N_0\equiv\frac{2\tau}{\tau_{\mathrm{des}}}\mathcal{N}_0.
\end{equation}
This trick will allows us to formulate the exchange equation also for site
$i=1$ in a homogeneous form. Namely, from Eq.~(\ref{exch}) we have
\begin{equation}
\label{exch2}
\frac{d\mathcal{N}_0(t)}{dt}=\frac{1}{2\tau}\left(N_1-N_0\right).
\end{equation}
Moreover, from Eqs.~(\ref{exch1}) we find
\begin{equation}
\label{exch3}
\frac{dN_1(t)}{dt}=\frac{1}{2\tau}\left(N_0-2N_1+N_2\right),
\end{equation}
and
\begin{equation}
\frac{dN_2(t)}{dt}=\frac{1}{2\tau}\left(N_1-2N_2+N_3\right),
\end{equation}
etc.

Let us now take the continuum limit. For that purpose we make a transition
from $\mathcal{N}_0\to n_s$ as the number of surface particles, and $N_i
\to an_b$ for the bulk concentration of particles. Expansion of the right
hand side of Eq.~(\ref{exch2}) yields the surface-bulk coupling
\begin{equation}
\frac{\partial n_s(t)}{\partial t}=\frac{a}{2\tau}a\left.\frac{\partial n_b
(\rho,t)}{\partial\rho}\right|_{\rho=0}.
\end{equation}
Similarly from Eq.~(\ref{exch3}) we obtain the bulk diffusion equation
\begin{equation}
\frac{\partial n_b(\rho,t)}{\partial t}=\frac{1}{2\tau}a^2\frac{\partial^2 n_b}{
\partial\rho^2}.
\end{equation}
Finally, the boundary condition
\begin{equation}
\left.\frac{a}{2\tau}n_b(\rho,t)\right|_{\rho=0}=\frac{1}{\tau_{\mathrm{des}}}
n_s
\end{equation}
stems from our definition (\ref{exch4}).

\section{Calculation of the surface mean squared displacement}
\label{msd_calc}

To calculate the quantity (\ref{surfmsd}) we start by rewriting expression
(\ref{nsol_fl}) in the form
\begin{equation}
n(k,s)=\frac{N_0}{s+k^2D_s+A(q,s)},
\end{equation}
where
\begin{equation}
q\equiv\sqrt{k^2+s/D_b}
\end{equation}
and
\begin{equation}
A(q,s)=\kappa q\frac{\Delta_1(q,s)}{\Delta(q,s)}.
\end{equation}
Differentiation of $n(k,s)$ yields
\begin{widetext}
\begin{equation}
\frac{\partial}{\partial k}n(k,s)=-\frac{N_0}{\left(s+k^2D_s+A(q,s)\right)^2}
\left(2kD_s+\frac{\partial A(q,s)}{\partial q}\frac{\partial q}{\partial k}
\right),
\end{equation}
and thus double differentiation of $n(k,s)$ produces
\begin{eqnarray}
\nonumber
\frac{\partial^2}{\partial k^2}n(k,s)&=&-\frac{N_0}{\left(s+k^2D_s+A(q,s)\right)
^2}\left(2D_s+\frac{\partial^2 A(q,s)}{\partial k^2}\right)\\
&&\hspace*{-0.8cm}
+\frac{2N_0}{\left(s+k^2D_s+A(q,s)\right)^3}\left(2kD_s+\frac{\partial A}{
\partial k}\right)^2.
\end{eqnarray}
\end{widetext}
Since
\begin{equation}
\left.\frac{\partial q}{\partial k}\right|_{k=0}=0,
\end{equation}
and therefore
\begin{equation}
\label{symm}
\left.\frac{\partial A(q,s)}{\partial k}\right|_{k=0}=0\,\,\,\mbox{and}\,\,\,
\left.\frac{\partial n(k,s)}{\partial k}\right|_{k=0}=0,
\end{equation}
such that the first moment $\langle z(t)\rangle$ of $n(z,t)$ vanishes, as it
should due to symmetry reasons. Moreover, we then obtain
\begin{eqnarray}
\nonumber
\langle z^2(s)\rangle&=&\frac{1}{\left[s+A(0,s)\right]^2}\left[2D_s+\left.
\frac{\partial^2A}{\partial k^2}\right|_{k=0}\right]\\
&=&\frac{N_s^2(s)}{N_0^2}\left[2D_s+\left.\frac{\partial^2A}{\partial k^2}
\right|_{k=0}\right].
\end{eqnarray}
The second term in the square brackets can be transformed to
\begin{equation}
\frac{\partial^2A(q,s)}{\partial k^2}=\frac{\partial}{\partial k}\left(
\frac{\partial A(q,s)}{\partial q}\right)\frac{\partial q}{\partial k}+\frac{
\partial A(q,s)}{\partial q}\frac{\partial^2q}{\partial k^2},
\end{equation}
such that
\begin{equation}
\left.\frac{\partial^2 A(q,s)}{\partial k^2}\right|_{k=0}=\sqrt{\frac{D_b}{s}}
\left.\frac{\partial A(q,s)}{\partial q}\right|_{k=0}.
\end{equation}
Differentiation of $A(q,s)$ results in the expression
\begin{equation}
\frac{\partial A(q,s)}{\partial q}=\kappa\frac{\Delta_1}{\Delta}\left[1+\frac{
q}{\Delta_1}\frac{\partial\Delta_1}{\partial q}-\frac{q}{\Delta}\frac{\partial
\Delta}{\partial q}\right].
\end{equation}
At $k=0$ we have $q=\sqrt{s/D_b}$, and we ultimately obtain
\begin{widetext}
\begin{equation}
\label{msd_gen}
\langle z^2(s)\rangle=\frac{N_s^2(s)}{N_0^2}\left\{2D_s+\kappa\sqrt{\frac{
D_b}{s}}\frac{\Delta_1(0,s)}{\Delta(0,s)}\left[1+\sqrt{\frac{s}{D_b}}
\left(\frac{1}{
\Delta_1(0,s)}\left.\frac{\partial\Delta_1}{\partial q}\right|_{k=0}-\frac{1}{
\Delta(0,s)}\left.\frac{\partial\Delta}{\partial q}\right|_{k=0}\right)\right]
\right\}.
\end{equation}
Here, we include the auxiliary quantities
\begin{equation}
\left.\frac{\partial\Delta_1}{\partial q}\right|_{k=0}=a\Big[K_1'\left(
\sqrt{st_a}\right)I_1\left(\sqrt{st_b}\right)-I_1'\left(\sqrt{st_a}\right)
K_1\left(\sqrt{st_b}\right)\Big]+b\Big[K_1\left(\sqrt{st_a}\right)I_1'
\left(\sqrt{st_b}\right)-I_1\left(\sqrt{st_a}\right)K_1'\left(\sqrt{st_b}
\right)\Big],
\end{equation}
and
\begin{equation}
\left.\frac{\partial\Delta}{\partial q}\right|_{k=0}=a\Big[I_1\left(
\sqrt{st_a}\right)K_1\left(\sqrt{st_b}\right)-K_1\left(\sqrt{st_a}\right)
I_1\left(\sqrt{st_b}\right)\Big]+b\Big[I_0\left(\sqrt{st_a}\right)K_1'
\left(\sqrt{st_b}\right)+K_0\left(\sqrt{st_a}\right)I_1'\left(\sqrt{st_b}
\right)\Big],
\end{equation}
\end{widetext}
In these expressions the prime implies a derivative over the whole argument,
and we have \cite{abramowitz}
\begin{equation}
\label{bess_diff}
I_1'(z)=I_0(z)-\frac{1}{z}I_1(z),\,\,\,\mbox{and}\,\,\,K_1'(z)=-K_0(z)-
\frac{1}{z}K_1(z).
\end{equation}

Taking the various limits corresponding to the cases discussed in Sections
\ref{sec_scl} and \ref{sec_icl} from above exact results we can indeed
confirm the results obtained there based on the limiting forms for the
surface propagator. We here demonstrate the corresponding derivations for
the surface mean squared displacement based on the exact Eq.~(\ref{msd_gen})
in the strong binding regime.

(i) At short times $t\ll t_{\kappa}\ll t_a\ll t_b$ we use the expansions
\cite{abramowitz}
\begin{equation}
\label{bess_exp}
I_{\nu}(z)\sim\frac{e^z}{\sqrt{2\pi z}},\,\,\,\mbox{and}\,\,\,
K_{\nu}(z)\sim\sqrt{\frac{\pi}{2z}}e^{-z}.
\end{equation}
Moreover, with Eq.~(\ref{bess_diff}) we find that
\begin{equation}
I_1'\left(\sqrt{st_{a/b}}\right)\sim I_0\left(\sqrt{st_{a/b}}\right)\sim
\frac{\exp\left(\sqrt{st_{a/b}}\right)}{(2\pi)^{1/2}\left(st_{a/b}\right)^{
1/4}}
\end{equation}
and
\begin{equation}
\label{m_bess_exp}
K_1'\left(\sqrt{st_{a/b}}\right)\sim-K_0\left(\sqrt{st_{a/b}}\right)\sim-
\frac{\pi^{1/2}\exp\left(-\sqrt{st_{a/b}}\right)}{2^{1/2}\left(st_{a/b}
\right)^{1/4}}
\end{equation}
so that in this approximation
\begin{equation}
\Delta_1(0,s)\sim\Delta(0,s)\sim\frac{\exp\left(s^{1/2}\left[t_b^{1/2}-t_a^{1/2}
\right]\right)}{2\left(st_a\right)^{1/4}\left(st_b\right)^{1/4}},
\end{equation}
thus
\begin{equation}
\left.\frac{\partial\Delta_1}{\partial q}\right|_{k=0}\sim\left.\frac{\partial
\Delta}{\partial q}\right|_{k=0}\sim(b-a)\frac{\exp\left(s^{1/2}\left[t_b^{1/2}
-t_a^{1/2}\right]\right)}{2(st_a)^{1/4}(st_b)^{1/4}},
\end{equation}
and finally
\begin{equation}
\label{bracket}
\left(\frac{1}{\Delta_1}\frac{\partial\Delta_1}{\partial q}-\frac{1}{\Delta}
\frac{\partial\Delta}{\partial q}\right)_{k=0}\sim0.
\end{equation}
From expressions (\ref{msd_gen}) and (\ref{N_lead}) we therefore obtain
\begin{equation}
\langle z^2(s)\rangle=\frac{2D_s}{s^2}+\frac{D_b}{s^{5/2}t_{\kappa}^{1/2}}.
\end{equation}
With the Laplace transform pair
\begin{equation}
\frac{1}{s^k}\div\frac{t^{k-1}}{\Gamma(k)},\,\,\,k>0,
\end{equation}
and $\Gamma(5/2)=3\pi^{1/2}/4$ we arrive at the result
\begin{equation}
\label{shortmsd}
\langle z^2(t)\rangle\sim2D_st
\left[1+\frac{2D_b}{3\pi^{1/2}D_s}\left(\frac{t}{t_{\kappa}}\right)^{1/2}
\right],
\end{equation}
where the equivalence with the normalized mean squared displacement is due to
the fact that $N_s(t)\sim N_0$ in this short time regime. This result
coincides with our finding (\ref{msd_res1}) obtained from the approximated
surface propagator.

(ii) At intermediate times $t_{\kappa}\ll t\ll t_a\ll t_b$ approximation
(\ref{bracket}) still holds. From Eq.~(\ref{msd_gen}) and with the time
evolution (\ref{num_int}) of the number of particle on the surface in this
regime, we obtain
\begin{eqnarray}
\nonumber
\langle z^2(s)\rangle&\sim&\frac{N_s(s)^2}{N_0^2}\left[2D_s+\kappa\left(
\frac{D_b}{s}\right)^{1/2}\right]\\
&\sim&2D_s\frac{t_{\kappa}}{s}+D_b\frac{t_{\kappa}^{1/2}}{s^{3/2}},
\end{eqnarray}
and, after inverse Laplace transformation,
\begin{equation}
\langle z^2(t)\rangle\sim2D_st_{\kappa}+\frac{2D_b}{\pi^{1/2}}t_{\kappa}^{1/2}
t^{1/2}.
\end{equation}
Again, we find coincidence with the previous result (\ref{msd_2}).

(iii) At even longer times $t_{\kappa}\ll t_a\ll t\ll t_b$ we make use of
approximations (\ref{bessapprox_4}) and (\ref{bess_exp}) to (\ref{m_bess_exp}).
For the derivatives of the Bessel functions we have in this regime
\begin{eqnarray}
\nonumber
I_1'\left(\sqrt{st_a}\right)&=&I_0\left(\sqrt{st_a}\right)-\frac{1}{\sqrt{
st_a}}I_1\left(\sqrt{st_a}\right)\sim\frac{1}{2},\\
\nonumber
K_1'\left(\sqrt{st_a}\right)&=&-K_0\left(\sqrt{st_a}\right)-\frac{1}{\sqrt{
st_a}}K_1\left(\sqrt{st_a}\right)\sim-\frac{1}{st_a},\\
\nonumber
I_1'\left(\sqrt{st_b}\right)&\sim&I_0\left(\sqrt{st_b}\right)\sim\frac{\exp
\left(\left[st_b\right]^{1/2}\right)}{(2\pi)^{1/2}(st_b)^{1/4}},\\
K_1'\left(\sqrt{st_b}\right)&\sim&-K_0\left(\sqrt{st_b}\right)\sim-\frac{
\pi^{1/2}\exp\left(-[st_b]^{1/2}\right)}{2^{1/2}(st_b)^{1/4}}.
\end{eqnarray}
Therefore,
\begin{equation}
\Delta_1(0,s)\sim\frac{\exp\left([st_b]^{1/2}\right)}{(2\pi)^{1/2}(st_a)^{1/2}
(st_b)^{1/4}}
\end{equation}
and
\begin{equation}
\left.\frac{\partial\Delta_1}{\partial q}\right|_{k=0}=\left(\frac{b}{(st_a)^{
1/2}}-\frac{a}{st_a}\right)\frac{b\exp\left((st_b)^{1/2}\right)}{(2\pi)^{1/2}
(st_b)^{1/4}},
\end{equation}
as well as
\begin{equation}
\Delta(0,s)\sim\frac{1}{2}\frac{\exp\left([st_b]^{1/2}\right)}{(2\pi)^{1/2}
(st_b)^{1/4}}\ln\left(\frac{4}{C^2st_a}\right)
\end{equation}
and
\begin{equation}
\left.\frac{\partial\Delta}{\partial q}\right|_{k=0}=\left[-\frac{a}{(st_a)^{
1/2}}+\frac{b}{2}\ln\left(\frac{4}{C^2st_a}\right)\right]\frac{\exp\left((st_b)
^{1/2}\right)}{(2\pi)^{1/2}(st_b)^{1/4}}.
\end{equation}
Here we define $C=\exp(\gamma)\approx1.78107$, where $\gamma\approx0.577216$ is
Euler's $\gamma$ constant. Collecting our results we find that
\begin{eqnarray}
\nonumber
&&\left.\frac{1}{\Delta_1(0,s)}\frac{\partial\Delta_1}{\partial q}\right|_{k=0}-
\left.\frac{1}{\Delta(0,s)}\frac{\partial\Delta}{\partial q}\right|_{k=0}\\
&&\hspace*{1.2cm}\sim
\frac{a}{(st_a)^{1/2}}\left(-1+\frac{2}{\displaystyle\ln\left(\frac{4}{C^2st_a}
\right)}\right)
\end{eqnarray}
and ultimately recover
\begin{equation}
\langle z^2(s)\rangle\sim\frac{N_s(s)^2}{N_0^2}\left[2D_s+\frac{4D_b}{(t_a
t_{\kappa})^{1/2}}\frac{1}{\displaystyle s\ln^2\left(\frac{4}{C^2st_a}\right)}
\right].
\end{equation}
Inserting expression (\ref{n_as}) we obtain
\begin{equation}
\langle z^2(s)\rangle\sim\frac{D_st_at_{\kappa}}{2}\ln^2\left(\frac{4}{C^2st_a}
\right)+\frac{D_b(t_at_{\kappa})^{1/2}}{s}.
\end{equation}
By Tauberian theorems the final result for the surface mean squared
displacements in this long time regime are
\begin{equation}
\langle z^2(t)\rangle\sim\frac{D_st_at_{\kappa}}{t}\ln\left(\frac{4t}{Ct_a}
\right)+D_b(t_at_{\kappa})^{1/2}.
\end{equation}
This result corroborates Eq.~(\ref{msd_3}).

(iv) Finally, at very long times $t_{\kappa}\ll t_a\ll t_b\ll t$
we have the asymptotic behaviors
\begin{eqnarray}
\nonumber
I_0\left(\sqrt{st_{a/b}}\right)&\sim&1,\\
\nonumber
K_0\left(\sqrt{st_{a/b}}\right)&\sim&-\gamma-\ln\left(\frac{(st_{a/b})^{1/2}}{2}
\right),\\
\nonumber
I_1\left(\sqrt{st_{a/b}}\right)&\sim&\frac{1}{2}(st_{a/b})^{1/2},\\
\nonumber
K_1\left(\sqrt{st_{a/b}}\right)&\sim&(st_{a/b})^{-1/2},\\
\nonumber
I_1'\left(\sqrt{st_{a/b}}\right)&=&I_0\left(\sqrt{st_{a/b}}\right)-\frac{
I_1\left(\sqrt{st_{a/b}}\right)}{(st_{a/b})^{1/2}}\sim\frac{1}{2},\\
\nonumber
K_1'\left(\sqrt{st_{a/b}}\right)&=&-K_0\left(\sqrt{st_{a/b}}\right)-
\frac{K_1\left(\sqrt{st_{a/b}}\right)}{(st_{a/b})^{1/2}}\\
&\sim&-\frac{1}{st_{a/b}}.
\end{eqnarray}
Therefore,
\begin{equation}
\left.\frac{\partial\Delta_1}{\partial q}\right|_{k=0}\sim\frac{b}{4}(st_a)^{1/2}
\ln\left(\frac{4}{C^2st_a}\right)
\end{equation}
and
\begin{equation}
\Delta_1(0,s)\sim\frac{1}{2}\sqrt{\frac{t_b}{t_a}}
\end{equation}
as well as
\begin{equation}
\left.\frac{\partial\Delta}{\partial q}\right|_{k=0}\sim-\frac{b}{st_b}+\frac{
b}{4}\ln\left(\frac{4}{C^2st_a}\right)
\end{equation}
and
\begin{equation}
\Delta(0,s)\sim\frac{1}{(st_b)^{1/2}}.
\end{equation}
For the square bracket in expression (\ref{msd_gen}) we obtain
\begin{equation}
\Big[1+\ldots\Big]\sim2,
\end{equation}
and thus
\begin{equation}
\langle z^2(s)\rangle\sim\frac{N_s^2(s)}{N_0^2}\left[2D_s+D_b\frac{t_b}{(t_a
t_{\kappa})^{1/2}}\right].
\end{equation}
With the help of Eq.~(\ref{ex_p}) this implies the result for the surface
mean squared displacement
\begin{equation}
\langle z^2(t)\rangle\sim\frac{8t_at_{\kappa}}{t_b^2}D_st+\frac{4(t_at_{
\kappa})^{1/2}}{t_b}D_bt.
\end{equation}
Thus, we corroborate the finding (\ref{msd_4}) from the approximate calculation.
With similar calculations one can reproduce the time evolution of the number
of particles on the inner cylinder surface and thus the normalized surface
mean squared displacement. Analogous reasoning confirms the results
obtained for the intermediate coupling limit in Sec.~\ref{sec_icl}.

\section{Calculation of the first passage time density}
\label{app_fpt}

In the Fourier-Laplace domain the diffusion equation (\ref{fpt_de}) becomes
an ordinary differential equation,
\begin{eqnarray}
\nonumber
&&\frac{d^2}{dr^2}P(r,k,s)+\frac{1}{r}\frac{d}{dr}P(r,k,s)-q^2P(r,k,s)\\
&&\hspace*{3.6cm}=-\frac{\delta(r-r_0)}{2\pi r_0D_b},
\label{de}
\end{eqnarray}
where
\begin{equation}
q^2=k^2+\frac{s}{D_b}.
\end{equation}
The boundary conditions are
\begin{equation}
P(r,k,s)\Big|_{r=a}=0
\end{equation}
and
\begin{equation}
\left.\frac{\partial}{\partial r}P(r,k,s)\right|_{r=b}=0.
\end{equation}

We rewrite the Fourier-Laplace transformed diffusion equation (\ref{de}) in
the form
\begin{equation}
\label{op_eq}
\mathbb{L}P(r,k,s)=\phi(r)
\end{equation}
where the operator $\mathbb{L}$ and the inhomogeneity $\phi(r)$ represent
\begin{equation}
\mathbb{L}=\frac{d^2}{dr^2}+\frac{1}{r}\frac{d}{dr}-q^2
\end{equation}
and
\begin{equation}
\phi(r)=\frac{\delta(r-r_0)}{2\pi r_0D_b}.
\end{equation}
Eq.~(\ref{op_eq}) can be solved by the method of variation of coefficients.
Namely, knowing that the solution of the homogeneous equation reads
\begin{equation}
P(r,k,s)=AI_0(qr)+BK_0(qr),
\end{equation}
to solve the full equation we assume that $A$ and $B$ are some functions of
the radius $r$. We impose the condition
\begin{equation}
A'(r)I_0(qr)+B'(r)K_0(qr)=0,
\end{equation}
where the prime denotes a derivative with respect to $r$. Consequently we
find
\begin{equation}
P'(r,k,s)=A(r)I_0'(qr)+B(r)K_0'(qr),
\end{equation}
and
\begin{eqnarray}
\nonumber
P''(r,k,s)&=&A(r)I_0''(qr)+B(r)K_0''(qr)\\
&&+A'(r)I_0'(qr)+B'(r)K_0'(qr).
\end{eqnarray}
According to the method of variation of coefficients this leads to
\begin{equation}
\mathbb{L}P(r,k,s)=A'(r)I_0'(qr)+B'(r)K_0'(qr).
\end{equation}
With above relations we arrive at a system of two equations with two unknowns,
$A'$ and $B'$,
\begin{eqnarray}
\nonumber
&&A'(r)I_0(qr)+B'(r)K_0(qr)=0\\
&&A'(r)I_0'(qr)+B'(r)K_0'(qr)=\phi(r),
\label{eqs}
\end{eqnarray}
The corresponding Wronskian is \cite{abramowitz}
\begin{eqnarray}
\nonumber
W(r)&=&I_0(qr)K_0'(qr)-K_0(qr)I_0'(qr)\\
\nonumber
&=&-q\Big(I_0(qr)K_1(qr)+I_1(qr)K_0(qr)\Big)\\
&=&-\frac{1}{r}.
\end{eqnarray}
The solutions to Eqs.~(\ref{eqs}) is
\begin{equation}
A'(r)=-\frac{1}{W(r)}K_0(qr)\phi(r)
\end{equation}
and
\begin{equation}
B'(r)=\frac{1}{W(r)}I_0(qr)\phi(r),
\end{equation}
and thus
\begin{eqnarray}
\nonumber
A(r)&=&-\int\frac{1}{W(r)}K_0(qr)\phi(r)dr\\
B(r)&=&\int\frac{1}{W(r)}I_0(qr)\phi(r)dr.
\end{eqnarray}
The general solution of Eq.~(\ref{op_eq}) has the form
\begin{eqnarray}
\nonumber
P(r,k,s)&=&\left(-\int\frac{1}{W(r)}K_0(qr)\phi(r)dr+A_0\right)I_0(qr)\\
&&\hspace*{-0.8cm}
+\left(\int\frac{1}{W(r)}I_0(qr)\phi(r)dr+B_0\right)K_0(qr),
\label{propa}
\end{eqnarray}
where the constants $A_0$ and $B_0$ are determined by the boundary
conditions that are rewritten as
\begin{widetext}
\begin{equation}
P(a,k,s)=\lim_{r\to a^+}\left\{\left(-\int_a^r\frac{1}{W(r')}K_0(qr')\phi(r')
dr'+A_0\right)I_0(qr)+\left(\int_a^r\frac{1}{W(r')}I_0(qr')\phi(r')dr'+B_0
\right)K_0(qr)\right\}=0
\label{bc1}
\end{equation}
and
\begin{equation}
\left.\frac{\partial}{\partial r}P(r,k,s)\right|_{r=b}=\left(-\int_a^b\frac{
1}{W(r')}K_0(qr')\phi(r')dr'+A_0\right)qI_0'(qb)+\left(\int_a^b\frac{1}{W(r')}
I_0(qr')\phi(r')dr'+B_0\right)qK_0'(qb)=0.
\label{bc2}
\end{equation}
\end{widetext}
Since the inhomogeneity $\phi(r)$ is a $\delta$ distribution, in general we
have to consider two cases separately: $r<r_0$ and $r>r_0$. However, as we
are interested solely in the first passage problem, for which we require the
probability flow at $r=a<r_0$ we restrict ourselves to the former case, only.
This means that Eq.~(\ref{bc1}) simplifies to
\begin{equation}
A_0I_0(qa)+B_0K_0(qa)=0.
\end{equation}
Performing the integrals in expression (\ref{bc2}) and using Eq.~(\ref{eqs})
we obtain the following system to determine the constants $A_0$ and $B_0$:
\begin{eqnarray}
\nonumber
&&A_0I_0(qa)+B_0K_0(qa)=0\\
\nonumber
&&A_0qI_1(qb)-B_0qK_1(qb)=\frac{q}{2\pi D_b}\Big(I_1(qb)K_0(qr_0)\\
&&\hspace*{4.2cm}+K_1(qb)I_0(qr_0)\Big)
\label{nileq}
\end{eqnarray}
with the following determinant
\begin{equation}
\mathrm{det}=-q\Big(I_0(qa)K_1(qb)+K_0(qa)I_1(qb)\Big).
\end{equation}
The solution of Eq.~(\ref{nileq}) is then
\begin{equation}
A_0=\frac{K_0(qa)}{2\pi D_b}\frac{I_1(qb)K_0(qr_0)+K_1(qb)I_0(qr_0)}{I_0(qa)
K_1(qb)+K_0(qa)I_1(qb)}
\end{equation}
and
\begin{equation}
B_0=-\frac{I_0(qa)}{2\pi D_b}\frac{I_1(qb)K_0(qr_0)+K_1(qb)I_0(qr_0)}{I_0(qa)
K_1(qb)+K_0(qa)I_1(qb)}.
\end{equation}
These two relations introduced into Eq.~(\ref{propa}) we obtain
\begin{widetext}
\begin{equation}
P(r,k,s)=\frac{1}{2\pi D_b}\frac{I_1(qb)K_0(qr_0)+K_1(qb)I_0(qr_0)}{I_0(qa)
K_1(qb)+K_0(qa)I_1(qb)}\Big(K_0(qa)I_0(qr)-I_0(qa)K_0(qr)\Big),\,\,\,r\le r_0.
\end{equation}
\end{widetext}
This result coincides with the findings of Berg and Blomberg \cite{berg},
when one takes the limit of a completely absorbing boundary condition [in
Berg and Blomberg's notation, this corresponds to $k\to\infty$ of their
reaction rate $k$].

\section{Some Laplace transforms}
\label{lap_trans}

We here provide a summary of non-trivial Laplace transforms involving
logarithmic functions, used throughout the text.

\subsection{Laplace inversion of the logarithm}

To calculate the inverse Laplace transforms in Eqs.~(\ref{n_as}) and
(\ref{nsurf_3}), as well as (\ref{formal1}) and (\ref{formal2}), we use
the direct Laplace transform of $\ln t$. Namely, for some $A>0$,
\begin{equation}
\mathscr{L}\left\{\ln\left(\frac{t}{A}\right)\right\}=\int_0^{\infty}e^{-st}
\ln\left(\frac{t}{A}\right)dt=-\frac{1}{s}\Big(\gamma+\ln(As)\Big),
\end{equation}
where $\gamma=-\int_0^{\infty}\exp(-x)\ln xdx\approx0.5772$ is Euler's constant.
Introducing the Laplace inversion with appropriate Bromwich path, we obtain
\begin{equation}
\int_{\mathrm{Br}}e^{st}\frac{\ln(As)}{s}\frac{ds}{2\pi i}=-\ln\left(\frac{Ct}{
A}\right),
\end{equation}
where $C=\exp(\gamma)$. Differentiation of this result with respect to $t$
yields
\begin{equation}
\label{lap1}
\int_{\mathrm{Br}}e^{st}\ln(As)\frac{ds}{2\pi i}=-\frac{1}{t},
\end{equation}
which is an exact result.

Eq.~(\ref{lap1}) delivers the desired result for Eq.~(\ref{nsurf_3}):
\begin{equation}
\mathscr{L}^{-1}\left\{\ln\left(\frac{4}{C^2st_a}\right)\right\}=\frac{1}{t},
\end{equation}
and for Eq.~(\ref{formal2}):
\begin{equation}
\mathscr{L}^{-1}\left\{\ln\Big(st_a+k^2a^2\Big)\right\}=-\frac{1}{t}e^{-k^2
a^2t/t_a}.
\end{equation}
For the latter relation we used the shift theorem of the Laplace transform:
\begin{equation}
\int_{\mathrm{Br}}e^{st}\ln(s+\Delta)\frac{ds}{2\pi i}=-\frac{1}{t}e^{\Delta t}.
\end{equation}

\subsection{Laplace inversion of the squared logarithm}

To obtain the Laplace pair (\ref{lap_pair}) we calculate
\begin{equation}
\int_{\mathrm{Br}}e^{st}\Big[\ln(As)\Big]^2\frac{ds}{2\pi i},\,\,\,A>0.
\end{equation}
We first deform the Bromwich path into the Hankel path $\mathrm{Ha}(\epsilon)$,
that is a loop starting from $-\infty$ below the negative real axis, merging
into a small circle around the origin with radius $|s|=\epsilon$, with $\epsilon
\to0$ in a positive sense, and finally receding to $-\infty$ moving above the
negative real axis. The integral along the Hankel path can be evaluated
directly:
\begin{widetext}
\begin{eqnarray}
\nonumber
I&=&\int_{\mathrm{Br}}e^{st}\Big[\ln(As)\Big]^2\frac{ds}{2\pi i}=\int_{\mathrm{
Ha}(\epsilon)}e^{st}\Big[\ln(As)\Big]^2\frac{ds}{2\pi i}\\
&=&\int_{-\infty}^0\exp\left(|s|te^{-i\pi}\right)\Big[\ln\left(A|s|e^{-i\pi}
\right)\Big]^2\frac{d|s|\exp(-i\pi)}{2\pi i}+\int_0^{\infty}\exp\left(|s|te^{
i\pi}\right)\Big[\ln\left(A|s|e^{i\pi}\right)\Big]^2\frac{d|s|\exp(i\pi)}{2\pi
i}+I_{\epsilon}.
\end{eqnarray}
\end{widetext}
Here $I_{\epsilon}\to0$ at $\epsilon\to0$ is the integral over the small circle
of radius $\epsilon$ around the origin. Collecting the terms we find
\begin{equation}
I=\int_0^{\infty}e^{-st}\Big[\ln(As)-i\pi\Big]^2\frac{ds}{2\pi i}+c.c=-2
\int_0^{\infty}e^{-st}\ln(As)ds,
\end{equation}
so that we finally obtain
\begin{equation}
\int_{\mathrm{Br}}e^{st}\Big[\ln(As)\Big]^2\frac{ds}{2\pi i}=\frac{2}{t}
\ln\left(\frac{Ct}{A}\right).
\end{equation}
Thus, we find Eq.~(\ref{lap_pair}),
\begin{equation}
\mathscr{L}^{-1}\left\{\ln^2\left(\frac{C^2t_a}{4}s\right)\right\}=\frac{2}{t}
\ln\left(\frac{4t}{Ct_a}\right).
\end{equation}

\subsection{Laplace inversion of the inverse square logarithm with additional
power}

We now address the Laplace pair of Eqs.~(\ref{aao}) and (\ref{aao1}), namely,
\begin{equation}
\mathscr{L}^{-1}\left\{s^{-3}\ln^{-2}\left(\frac{4}{C^2st_a}\right)\right\}=
\frac{t^2}{2}\ln^{-2}\left(\frac{4t}{C^2t_a}\right).
\end{equation}
To see this result let us recall the Tauberian theorems, see, for instance,
Ref.~\cite{feller}. These state that if the Laplace transform of some (positive)
function $\omega(t)$ behaves like
\begin{equation}
\omega(s)\sim s^{-\rho}L\left(\frac{1}{s}\right),\,\,\,s\to0,\,\,\,0\le\rho<
\infty,
\end{equation}
then its inverse Laplace transform has the asymptotic form
\begin{equation}
\omega(t)\sim\frac{t^{\rho-1}}{\Gamma(\rho)}L(t),\,\,\, t\to\infty.
\end{equation}
Here $L(t)$ is a function slowly varying at infinity, i.e.,
\begin{equation}
\lim_{x\to\infty}\frac{L(ax)}{L(x)}=1\,\,\,\forall\,\,\,a>0.
\end{equation}

\end{appendix}


\begin{thebibliography}{99}
\bibitem{REMMM} In our problem we deal with a truncated Cauchy distribution,
therefore the normalization factor in Eq.~(\ref{cauchy}) is actually
different from $1/\pi$. However, as the truncation is accomplished by the
rapidly decaying compressed Gaussian, the correction to $1/\pi$ is expected
to be small, and the Cauchy part contains almost the entire probability,
\[
\frac{2}{\pi}\int_0^{\sqrt{D_bt}}\frac{\kappa t}{z^2+\kappa^2 t^2}dz=
\frac{2}{\pi}\arctan\sqrt{\frac{t_{\kappa}}{t}}\approx1,
\]
since $t\ll t_{\kappa}$.
\end{thebibliography}
\end{document}